\documentclass[pra,showpacs,twocolumn,aps,superscriptaddress,a4paper,longbibliography]{revtex4-1}
\usepackage{dcolumn,amssymb,amsmath,amsfonts,graphicx,latexsym,color}
\usepackage[utf8]{inputenc}
\usepackage[T1]{fontenc}
\usepackage{epstopdf}

\usepackage{ulem}
\usepackage{epstopdf}
\usepackage{subfigure}
\usepackage{float}
\usepackage{mathrsfs}
\usepackage{bbm}
\usepackage{psfrag}
\usepackage{CJK}

\usepackage{verbatim}
\usepackage{multirow}
\usepackage{diagbox}
\usepackage[colorlinks,citecolor=blue]{hyperref}
\usepackage{amsmath,amssymb,amsfonts,bm}
\usepackage{graphicx}
\usepackage{epstopdf}
\usepackage{dcolumn}
\usepackage{mathrsfs}
\usepackage{bbold}
\usepackage{dsfont}
\usepackage{dcolumn}

\begin{document}

\title{Emergent Weyl-like points in periodically modulated systems}

\author{Fang Qin}
\email{qinfang@just.edu.cn}
\affiliation{School of Science, Jiangsu University of Science and Technology, Zhenjiang, Jiangsu 212100, China}

\author{Rui Chen}
\email{chenr@hubu.edu.cn}
\affiliation{Department of Physics, Hubei University, Wuhan, Hubei 430062, China}

\begin{abstract}
We investigate a three-dimensional (3D) topological phase resembling a Weyl semimetal, modulated by a periodic potential and engineered through Floquet dynamics. This system is constructed by stacking two-dimensional Chern insulators and hosts Weyl-like points defined in the parameter space $(k_x, k_y, z)$, distinct from conventional Weyl points in momentum space $(k_x, k_y, k_z)$. The Weyl-semimetal-like phase exhibits characteristics akin to those of Weyl semimetals, including linear dispersion near the Weyl-like points, nontrivial bulk topology, the presence of Fermi arcs connecting the Weyl-like points, and the Berry monopoles. Unlike traditional Weyl semimetals, these features manifest in real space rather than momentum space. Furthermore, we calculate the local density of states, the layer Hall conductance, and the total 3D Hall conductivity, demonstrating that the Weyl-semimetal-like phase remains stable under weak and moderate interlayer couplings. The influence of disorder is also examined: Beyond a critical disorder strength, the Weyl-like points destabilize and the topological phase collapses. Moreover, by computing the Floquet Chern number, we demonstrate that the locations of the Weyl-like points can be tuned via high-frequency laser pumping. Finally, we show that both type-I and II Weyl-like behaviors can arise in a tilted Weyl-like model.
\end{abstract}

\maketitle

\section{Introduction}

Stacks of two-dimensional (2D) Chern insulator phases have proven to be a versatile platform for realizing extraordinary topological phases, attracting considerable attention in recent years~\cite{mogi2017tailoring,mogi2017magnetic,xiao2018realization,mogi2022experimental,qin2023light,chen2023half,chang2013experimental,yoshimi2015quantum,shan2010effective,lu2010massive,li2010chern,lu2013quantum,chen2019effects,sun2020analytical,liu2010model,dabiri2021light,dabiri2021engineering,sun2023magnetic,qin2022phase,zhu2023floquet}. For instance, the well-known Weyl semimetal phase can be interpreted as a stack of 2D Chern insulators with orbital-dependent interlayer tunneling~\cite{yang2011quantum,chen2015disorder,chen2018floquet2,chen2018phase,li2021type}. Similarly, phases such as the axion insulator and the hinged quantum spin Hall effect arise from alternating stacks of Chern insulators with opposite Chern numbers~\cite{chen2021field,chen2022layer,chen2023side,chen2023chiral}. Moreover, higher-Chern-number quantum anomalous Hall insulators can be constructed by alternating stacks of Chern insulators and normal insulators~\cite{zhao2022zero,baba2022effect,li2022chern,li2023quantum}.

Weyl semimetals are characterized by linearly dispersing band-touching points, known as Weyl points, in momentum space~\cite{wan2011topological,chan2016chiral,yan2017topological,guo2023light}. These Weyl points are topologically protected and act as monopoles or antimonopoles of the Berry curvature~\cite{li2016negative,lu2017quantum}. According to the no-go theorem~\cite{nielsen1981absence1,nielsen1981absence2}, Weyl points always appear in pairs and are connected by surface states known as Fermi arcs.
Recent advancements have realized ideal Weyl semimetals across various platforms. They have been implemented in electrical circuits~\cite{lu2019probing,lee2020imaging}, ultracold atomic gases with three-dimensional (3D) spin-orbit coupling~\cite{dubvcek2015weyl,roy2018tunable,wang2021realization,lu2020ideal}, photonic crystals~\cite{lu2015experimental,li2018weyl}, and acoustic metamaterials~\cite{xiao2015synthetic,yang2016acoustic,peri2019axial}.

Zhang {\emph{et al.}}~\cite{zhang2023emergent} recently proposed and studied a 2D phase modulated by a periodic potential. Unlike conventional Chern insulators, where the spectral flow of edge modes is momentum dependent, their system exhibits edge modes whose spectral flow depends on spatial position. They also suggested that such a periodic potential could be experimentally realized via charge density wave effects~\cite{huang2021absence,qin2020theory}. In a related development, Chen {\emph{et al.}}~\cite{chen2024tunable} explored a 2D higher-order topological insulator and demonstrated that the energies of the corner modes can be tuned by a periodic spin-orbital potential. Additionally, it has been shown that corner modes can also be controlled via long-range hoppings~\cite{qin2025tunable}.

Off-resonant light serves as a powerful tool for inducing Floquet Weyl semimetals~\cite{wang2014floquet, hubener2017creating}. Building on Floquet theory~\cite{oka2009photovoltaic,zhan2024perspective,qin2023light,qin2022light,qin2022phase,qin2024light,dabiri2021light,dabiri2021engineering,dabiri2022floquet,askarpour2022light,pervishko2018impact,calvo2015floquet,seshadri2022engineering,seshadri2019generating,seshadri2022floquet,lee2018floquet,magnus1954exponential,blanes2009magnus,bukov2015universal,eckardt2015high,chen2018floquet1,chen2018floquet2,du2022weyl,wang2022floquet}, it would be intriguing to investigate the experimental implications of realistic periodic driving frequencies. This is particularly relevant in light of recent developments, such as the photoinduced anomalous Hall effect in graphene~\cite{mciver2020light} and its corresponding microscopic theory~\cite{sato2019microscopic}, the experimental observation of Floquet states in graphene~\cite{merboldt2024observation}, and the direct detection of Floquet-Bloch states in monolayer graphene~\cite{choi2024direct}, WSe$_2$~\cite{aeschlimann2021survival}, and the 3D topological insulator Bi$_2$Se$_3$~\cite{wang2013observation}. Additionally, authors of experimental studies on Floquet transport effects, including anomalous linear transport in light-driven 3D Dirac electrons in bismuth~\cite{hirai2023anomalous} and nonperturbative nonlinear transport in a magnetic Floquet-Weyl semimetal~\cite{day2024nonperturbative}, have further highlighted the potential of this approach.

In this work, we propose and study a 3D topological phase resembling a Weyl semimetal, modulated by a periodic potential and realized through a simple stack of 2D Chern insulators. The system features Weyl-like points defined in the parameter space $(k_x, k_y, z)$, distinct from the conventional Weyl points in momentum space $(k_x, k_y, k_z)$. This Weyl-semimetal-like phase exhibits key properties of a Weyl semimetal, including linear dispersion near the Weyl-like points, nontrivial bulk topology, the emergence of Fermi arc connecting these points, and the Berry monopoles. However, these behaviors are observed in real space rather than momentum space, as in traditional Weyl semimetals. We further compute the local density of states, the layer Hall conductance, and the total 3D Hall conductivity, demonstrating that the Weyl-semimetal-like phase remains robust under weak and moderate interlayer couplings. The impact of disorder is also analyzed: Beyond a critical disorder strength, the Weyl-like points destabilize and the topological phase collapses. In addition, we evaluate the Floquet Chern number under high-frequency driving and show that the positions of the Weyl-like nodes can be effectively tuned by laser fields. Finally, we reveal the emergence of both type-I and II Weyl-like behaviors in a tilted Weyl-like model.

The structure of this paper is as follows. In Section~\ref{2}, we introduce the Weyl-like model Hamiltonian with a periodic potential along the $z$-direction, alongside the conventional Weyl semimetal model, and discuss the duality between these two models. Section~\ref{3} presents calculations of the local density of states, the layer Hall conductances, and the total 3D Hall conductivity in the presence of interlayer hopping for the Weyl-like model. 
In Section~\ref{4}, we investigate the effects of a generic disorder potential on the Weyl-like points and the associated topological invariants.
In Section~\ref{5}, we derive the Floquet effective Hamiltonians for both models in Section~\ref{2} under high-frequency laser driving. Section~\ref{6} details the analysis of Floquet Weyl nodes, Floquet Chern numbers, the Floquet energy bands, and Floquet Fermi arcs for both models. A brief discussion of possible experimental realizations of the Weyl-like model is given in Section~\ref{7}.
In Section~\ref{8}, we investigate the realization for both type-I and II Weyl-like model behaviors in a tilted Weyl-like model. Finally, Section~\ref{9} summarizes the main results of the paper.

\section{Model}\label{2}

In this section, we will introduce the Weyl-like and Weyl semimetal models.

\subsection{Weyl-like model}\label{2.1}

We begin by introducing a two-band model Hamiltonian that describes a stack of 2D Chern insulators subjected to a periodic potential along the $z$ direction. This model is defined in the $(k_x, k_y, z)$ space as
\begin{eqnarray}
{\cal H}({\bf k}_{||},z)&\!=\!&h_{\text{2D}}({\bf k}_{||})\!+\!V(z)\sigma_{z}, \label{eq:H_z}
\end{eqnarray}
where $h_{\text{2D}}({\bf k}_{||})\!=\!m_{z}\sigma_{z}\!+\!m_{0}[2\!-\!\cos(k_{x}a)\!-\!\cos(k_{y}a)]\sigma_z\!+\!t_{x}\sin(k_{x}a)\sigma_{x}\!+\!t_{y}\sin(k_{y}a)\sigma_{y}$ represents the Hamiltonian of a 2D Chern insulator when $m_{z}\!\times\! m_{0}\!<\!0$~\cite{shen2017topological}, and ${\bf k}_{||}\!=\!(k_x, k_y)$. Here, $a$ is the lattice constant, and $\sigma_{i}$ $(i\!=\!x,y,z)$ are Pauli matrices acting on the orbital space. The spatially periodic potential is given by $V(z)\!=\!V_{z}\cos(Q_{z}z)$, where $V_{z}$ is the potential strength, $z\!=\!j_{z}a$ denotes the position of the $j_z$th layer along the $z$ direction, and $\lambda_{z} \!=\! 2\pi/Q_{z}$ is the wavelength of the modulation.

We emphasize that the layers are decoupled in this case, ensuring that $z$ remains a good quantum number. The effects of interlayer coupling will be addressed in Section~\ref{3}.

\begin{figure}[htpb]
\centering
\includegraphics[width=\columnwidth]{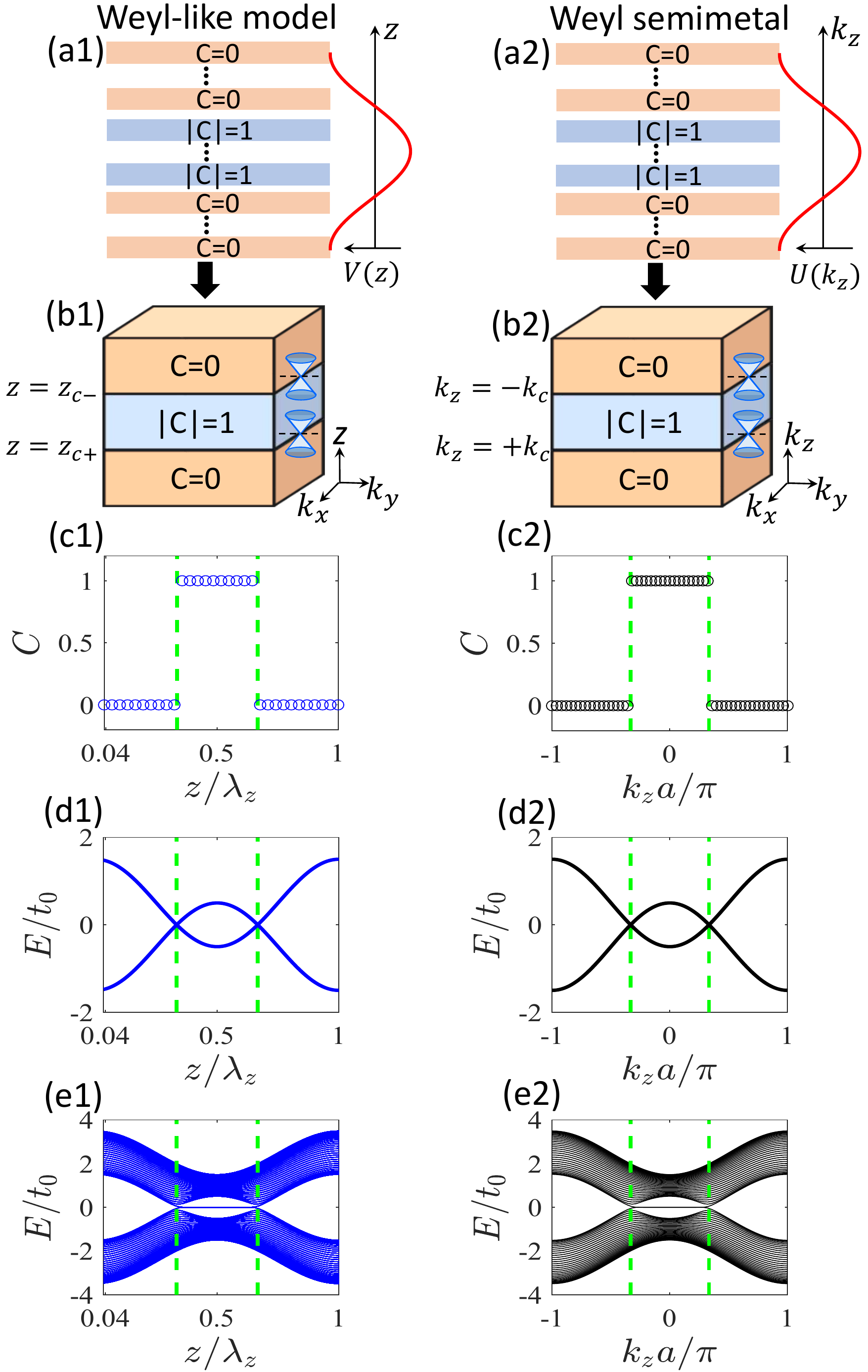}
\caption{Duality between the decoupled Weyl-like model (left column) and the Weyl semimetal model (right column). The two green vertical dashed lines indicate the positions of the two Weyl-like (or Weyl) points. (a1) Schematic of the decoupled Weyl-like model in real space, featuring a spatially periodic potential $V(z)\!=\!V_{z}\cos(Q_z z)$, where $Q_{z}\!=\!2 \pi/\lambda_z$. (b1) Weyl-like cones in the $(k_x,k_y,z)$ space at two critical planes, $(k_x,k_y,z_{c-})$ and $(k_x,k_y,z_{c+})$, marking topological transitions. (c1) Chern number of each layer [Eq.~\eqref{eq:Chern_z}] as a function of $z$ for the decoupled Weyl-like model with $V_{z}\!=\!1$ eV, $\lambda_z\!=\!L_{z}\!=\!31a$, and the lattice constant $a\!=\!1$ nm. (d1) Energy bands [Eq.~\eqref{eq:E_z}] (blue solid curves) of the Weyl-like model at $k_{x}\!=\!k_{y}\!=\!0$. (e1) Energy bands (blue solid curves) and Fermi-like arc (blue horizontal line for $E\!=\!0$) for the Weyl-like model under open boundary conditions (OBCs) along the $y$ direction, with $k_x\!=\!0$ and system size $L_y\!=\!31a$. (a2) Schematic of the Weyl semimetal model in real space, featuring a momentum-dependent potential $U(k_z)\!=\!t_{z}\cos(k_{z}a)$. (b2) Weyl cones in momentum space $(k_x,k_y,k_z)$ at two critical planes $(k_x,k_y,k_{c-})$ and $(k_x,k_y,k_{c+})$, indicating the topological features of the model. (c2) Chern number [Eq.~\eqref{eq:Chern_Weyl}] as a function of $k_z$ for the Weyl semimetal model with $t_{z}\!=\!1$ eV. (d2) Energy bands [Eq.~\eqref{eq:E_Weyl}] (black solid curves) of the Weyl semimetal model at $k_{x}\!=\!k_{y}\!=\!0$. (e2) Energy bands (black solid curves) and Fermi arc (black horizontal line at $E\!=\!0$) for the Weyl semimetal model under OBCs along the $y$ direction, with $k_x\!=\!0$ and system size $L_{y}\!=\!31a$. Other parameters used: $m_{z}\!=\!0.5$ eV, $m_{0}\!=\!1$ eV, $t_{x}\!=\!t_{y}\!=\!t_{0}\!=\!1$ eV, and $L_{z}\!=\!31a$.}
\label{fig_dual}
\end{figure}

As illustrated in Fig.~\ref{fig_dual}(a1), the system described by the Weyl-like Hamiltonian in Eq.~\eqref{eq:H_z} can be interpreted as an array of decoupled 2D layers, each with a spatially dependent effective mass term $\tilde{m}_z(z)\!=\!m_z\!+\!V(z)$. For the calculations presented in this work, we adopt the fixed parameters $m_{z}\!=\!0.5$ eV, $m_{0}\!=\!1$ eV, $t_x\!=\!t_y\!=\!t_0\!=\!1$ eV, $V_{z}\!=\!1$ eV, and $\lambda_{z}\!=\!31a$, with $a\!=\!1$ nm. The detailed matrix representation of the real-space Hamiltonian with the periodic potential along the $z$ direction, assuming open boundary conditions (OBCs) along the $y$ direction and periodic boundary conditions (PBCs) along the $x$ direction, is provided in Section SI in the Supplemental Material~\cite{SuppMat}.

The eigenenergies of the Weyl-like model described by Eq.~\eqref{eq:H_z} are given by
\begin{eqnarray}
E_{\pm}\!=\!\pm\sqrt{[t_{x}\sin (k_{x}a)]^{2}\!+\![t_{y}\sin (k_{y}a)]^{2}\!+\![\tilde{M}_{z}(k_{x},k_{y},z)]^{2}},\label{eq:E_z0}\nonumber\\
\end{eqnarray} where $\tilde{M}_{z}(k_{x},k_{y},z)\!=\!\tilde{m}_{z}(z)\!+\!m_{0}[ 2\!-\!\cos (k_{x}a)\!-\!\cos (k_{y}a)]$.

Since the layers are decoupled from each other, we can define a $z$-dependent Chern number of each layer for the $m$th occupied band as~\cite{fukui2005chern,sun2020analytical,otrokov2019unique,gao2021layer,chen2022layer,mong2010antiferromagnetic,essin2009magnetoelectric,wang2015quantized,varnava2018surfaces,chen2023half,fu2021bulk,price2015four,zhang2019entangled}
\begin{eqnarray}
C_{m}(z)\!=\!\frac{1}{2\pi}\int{\cal F}_{xy}^{mm}({\bf k}_{||},z)d^{2}{\bf k}_{||},\label{eq:Chern_z}
\end{eqnarray} where the non-Abelian Berry curvature is defined as~\cite{chen2023half,fu2021bulk,price2015four,zhang2019entangled}
\begin{eqnarray}
{\cal F}_{xy}^{mn}({\bf k}_{||},z)&\!=\!&\partial_{k_x}{\cal A}_{k_y}^{mn}({\bf k}_{||},z)-\partial_{k_y}{\cal A}_{k_x}^{mn}({\bf k}_{||},z) \nonumber\\
&&\!+ i[{\cal A}_{k_x}^{mn}({\bf k}_{||},z),{\cal A}_{k_y}^{mn}({\bf k}_{||},z)],\label{eq:Berry_z}
\end{eqnarray} and the Berry connection ${\cal A}_{k_\alpha}^{mn}({\bf k}_{||},z)$ of the occupied bands (with $m,n\in{\rm occ}$) is given by
\begin{eqnarray}
{\cal A}_{k_\alpha}^{mn}({\bf k}_{||},z)\!=\!i\langle\psi_{m}|\partial_{k_\alpha}\psi_{n}\rangle.\label{eq:Berry_connection_z}
\end{eqnarray} 
Here, ``occ\rq{}\rq{} denotes the set of occupied bands. A detailed derivation of the non-Abelian Berry curvature in Eq.~\eqref{eq:Berry_z} is provided in Section SII in the Supplemental Material~\cite{SuppMat}.

The $z$-dependent Chern number of each layer is illustrated in Fig.~\ref{fig_dual}(c1).
Figures~\ref{fig_dual}(a1) and \ref{fig_dual}(b1) show that each layer behaves as a Chern insulator with Chern number $|C|\!=\!1$ for $z_{c-}\!<\!z\!<\!z_{c+}$ and as a normal insulator with $C\!=\!0$ outside this region (i.e., for $z\!<\!z_{c-}$ and $z\!>\!z_{c+}$). To determine the critical values $z_{c\pm}$, we evaluate the eigenenergies of the Weyl-like Hamiltonian in Eq.~\eqref{eq:H_z} at the high-symmetry point $k_x\!=\!k_y\!=\!0$, obtaining from Eq.~\eqref{eq:E_z0}
\begin{eqnarray}
E_{\pm}\!=\!\pm\tilde{m}_z(z)\!=\!\pm\left[ m_{z}\!+\!V_{z}\cos(Q_{z}z)\right].\label{eq:E_z}
\end{eqnarray} Furthermore, the critical values $z_{c\pm}$ correspond to the gap-closing points, where $E_{\pm}\!=\!\pm\tilde{m}_z(z)\!=\!\pm\left[ m_{z}\!+\!V_{z}\cos(Q_{z}z)\right]\!=\!0$, leading to
\begin{eqnarray}
z_{c\pm}\!=\!\frac{\lambda_{z}}{2\pi}\left[\pi\pm\arccos\left(\frac{m_{z}}{V_{z}}\right)\right],\label{eq:z_c}
\end{eqnarray} under the condition $-1\!\leqslant\! m_{z}/V_{z}\!\leqslant\!1$.

Figure~\ref{fig_dual}(d1) shows the energy band as a function of position $z$ at $k_x\!=\!k_y\!=\!0$. A clear band inversion occurs near $z\!=\!z_{c\pm}$, highlighting the spectral flow characteristic of the layered topological structure.
Additionally, Fig.~\ref{fig_dual}(e1) reveals the presence of a Fermi arc, further demonstrating the topological nature of the system.

Moreover, the total 3D Hall conductivity for the $m$th occupied band is given by~\cite{burkov2011weyl}
\begin{eqnarray}
\sigma_{H}^{3D}&\!=\!&\sum_{j_z=0}^{L_{z}/a}\frac{\sigma_{H}^{2D}(z)}{L_{z}}
\!=\!\frac{e^{2}}{h}\!\sum_{j_z=0}^{L_{z}/a}\!\frac{C_{m}(z)}{L_{z}},  \label{eq:conductivity3D_Weyl_like}
\end{eqnarray} where $j_{z}\!=\!z/a$ is the layer index, $L_z$ is the system length along the $z$ direction, $\sigma_{H}^{2D}(z)\!=\!(e^{2}/h)C_{m}(z)$ is the Hall conductance of each layer associated with the $m$th occupied band, $C_{m}(z)$ is defined in Eq.~\eqref{eq:Chern_z}, and $L_{z}$ denotes the system size along the $z$ direction. 
Using the critical values $z_{c\pm}$ obtained in Eq.~\eqref{eq:z_c}, we can simplify the 3D Hall conductivity expression as
\begin{eqnarray}
\sigma_{H}^{3D}
&\!=\!&\frac{e^{2}}{h}\sum_{j_z=z_{c-}/a}^{z_{c+}/a}\frac{C_{m}(z)}{L_{z}} \nonumber\\
&\!\approx\!& \!\frac{e^{2}\lambda_{z}}{hL_{z}a\pi}\arccos\left(\frac{m_{z}}{V_{z}}\right)\!, \label{eq:conductivity3D_last}
\end{eqnarray}
where the approximation in the second line assumes $L_{z}\!\gg\! a$, allowing the sum to be replaced by an integral over 
$z$. This result highlights how the 3D Hall response is governed by the range of layers exhibiting nontrivial topology.

The specific choice of the finite system length $L_z$ along the $z$ direction plays a crucial role in the topological analysis. In our numerical calculations, we take a finite system length $L_{z}\!=\!\lambda_{z}$ along the $z$ direction, where $\lambda_{z}$ is the modulation period such that $V(z)\!=\!V(z\!+\!\lambda_{z})$. Under this choice, two Weyl-like points appear along the $z$ direction in the finite system. 

If a shorter length is chosen, for example, $L_{z}\!<\!z_{c-}$, no Weyl-like point exists in the finite system. If $z_{c-}\!<\!L_{z}\!<\!z_{c+}$, there will be only one Weyl-like point at $z_{c-}$. Conversely, for a longer length, such as $L_{z}\!=\!2\lambda_{z}$, four Weyl-like points will emerge in the finite system.

Therefore, although $V(z)$ is spatially periodic along $z$, the way this periodicity enters the topological analysis depends on the specific choice of the finite system length $L_{z}$ in the $z$ direction.

\subsection{Weyl semimetals}\label{2.2}

To facilitate a comparative study, we now review the Weyl semimetal model, whose Hamiltonian is given by~\cite{yang2011quantum,chen2015disorder,chen2018floquet2,chen2018phase}
\begin{eqnarray}
{\cal H}_{\rm W}({\bf k})&\!=\!&\left[-m_{z}\!+\!t_{z}\cos(k_{z}a)\right]\sigma_{z} \nonumber\\
&&\!-m_{0}\left[2\!-\!\cos(k_{x}a)\!-\!\cos(k_{y}a)\right]\sigma_{z} \nonumber\\
&&\!+t_{x}\sin(k_{x}a)\sigma_{x}\!+\!t_{y}\sin(k_{y}a)\sigma_{y},\label{eq:H_Weyl}
\end{eqnarray}
where ${\bf k}\!=\!(k_x,k_y,k_z)$. This Hamiltonian describes a Weyl semimetal phase and can be viewed as a stack of 2D layers with orbital-dependent interlayer hopping $t_z$, as illustrated in Fig.~\ref{fig_dual}(a2). 
The matrix representation of Eq.~\eqref{eq:H_Weyl} in real space with OBCs along the 
$z$ direction and PBCs along the $x$ and $y$ directions is provided in Subsection SIII A in the Supplemental Material~\cite{SuppMat}. Similarly, the matrix form with OBCs in the $y$ direction and PBCs in the $x$ and $z$ directions is given in Subsection SIII B in the Supplemental Material~\cite{SuppMat}.

We can rewrite Eq.~\eqref{eq:H_Weyl} as ${\cal H}_{\rm W}({\bf k})\!=\!\mathbf{h}({\bf k})\cdot{\pmb\sigma}$,  where the vector $\mathbf{h}({\bf k})\!=\!(h_{x},h_{y},h_{z})$ has the following components: $h_{x}\!=\!t_{x}\sin(k_{x}a)$, $h_{y}\!=\!t_{y}\sin(k_{y}a)$, and $h_{z}\!=\!M_{z}(k_z)\!-\!m_{0}\left[2\!-\!\cos(k_{x}a)\!-\!\cos(k_{y}a)\right]$, with $M_{z}(k_{z})\!=\!-m_{z}\!+\!t_{z}\cos(k_{z}a)$. The eigenenergies of the Weyl semimetal model then take the form
\begin{eqnarray}
E_{{\rm W},\pm}\!=\!\pm\sqrt{h_{x}^{2}\!+\!h_{y}^{2}+h_{z}^{2}}. \label{eq:E_Weyl0}
\end{eqnarray}

The Chern number of this model in Eq.~\eqref{eq:H_Weyl} at fixed $k_{z}$ can be calculated using the expression~\cite{thouless1982quantized,fukui2005chern,xiong2016towards,liu2019floquet,lu2010massive,shan2010effective,qin2022phase}
\begin{eqnarray}
C(k_z)\!=\!\frac{1}{4\pi}\int_\text{BZ}\!\!d^{2}{\bf k}_{||}\frac{\mathbf{h}({\bf k})}{|\mathbf{h}({\bf k})|^{3}}\!\cdot\![\partial_{k_{x}}\mathbf{h}({\bf k})\!\times\!\partial_{k_{y}}\mathbf{h}({\bf k})],\label{eq:Chern_Weyl}\nonumber\\
\end{eqnarray} where the integral is over the 2D Brillouin zone (BZ). A detailed analytical expression for the Chern number in Eq.~\eqref{eq:Chern_Weyl} is provided in Section SIV in the Supplemental Material~\cite{SuppMat}. 

For the special case $k_x\!=\!k_y\!=\!0$, the eigenenergies simplify to
\begin{eqnarray}
E_{\pm}\!=\!\pm\left[-m_{z}\!+\!t_{z}\cos(k_{z}a)\right], \label{eq:E_Weyl}
\end{eqnarray} so the corresponding bulk energy gap at $k_x\!=\!k_y\!=\!0$ is given by $\Delta_{\rm W}\!=\!E_{{\rm W},+}\!-\!E_{{\rm W},-}\!=\!2|M_{z}(k_z)|$, as shown in Fig.~\ref{fig_dual}(d2).
In this case, the system hosts a pair of gapless Weyl nodes at $(0,0,k_{c\pm})$ with the positions $k_{c\pm}\!=\!\pm(1/a)\arccos(m_{z}/t_{z})$, under the condition $|m_{z}/t_{z}|\leqslant1$, as shown in Figs.~\ref{fig_dual}(b2) and \ref{fig_dual}(c2). These Weyl points $k_{c\pm}$ are determined by solving the condition $M_{z}(k_{z})\!=\!-m_{z}\!+\!t_{z}\cos(k_{z}a)\!=\!0$.
For $k_z$ in the range $k_{c-}\!<\!k_{z}\!<\!k_{c+}$, the system behaves as a Chern insulator with Chern number $|C|\!=\!1$. Outside this range, for $k_{z}\!<\!k_{c-}$ or $k_{z}\!>\!k_{c+}$, the system becomes a normal insulator with $C\!=\!0$, as depicted in Fig.~\ref{fig_dual}(c2).
The corresponding Fermi arc is shown in Fig.~\ref{fig_dual}(e2).

Finally, the total 3D Hall conductivity for the Weyl semimetal model in Eq.~\eqref{eq:H_Weyl} is given by~\cite{burkov2011weyl}
\begin{eqnarray}
\sigma_{H}^{3D}\!=\!\int_{-\pi/a}^{\pi/a}\frac{dk_{z}}{2\pi}\sigma_{H}^{2D}(k_z)\!=\!\frac{e^{2}}{h}\int_{-\pi/a}^{\pi/a}\frac{dk_{z}}{2\pi}C(k_z), \label{eq:conductivity3D_Weyl}
\end{eqnarray} where $\sigma_{H}^{2D}(k_z)\!=\!(e^{2}/h)C(k_z)$ is the $k_{z}$-resolved 2D Hall conductance and $C(k_z)$ is defined in Eq.~\eqref{eq:Chern_Weyl}.
Using the Weyl node positions $k_{c\pm}\!=\!\pm(1/a)\arccos(m_{z}/t_{z})$, we obtain
\begin{eqnarray}
\sigma_{H}^{3D}
\!=\!\frac{e^{2}}{h}\int_{k_{c-}}^{k_{c+}}\frac{dk_{z}}{2\pi}C(k_z)\!=\!\frac{e^{2}}{ha\pi}\arccos\left(\!\frac{m_{z}}{t_{z}}\!\right)\!. \label{eq:conductivity3D_Weyl_last}
\end{eqnarray}

\begin{figure}
\centering
\includegraphics[width=\columnwidth]{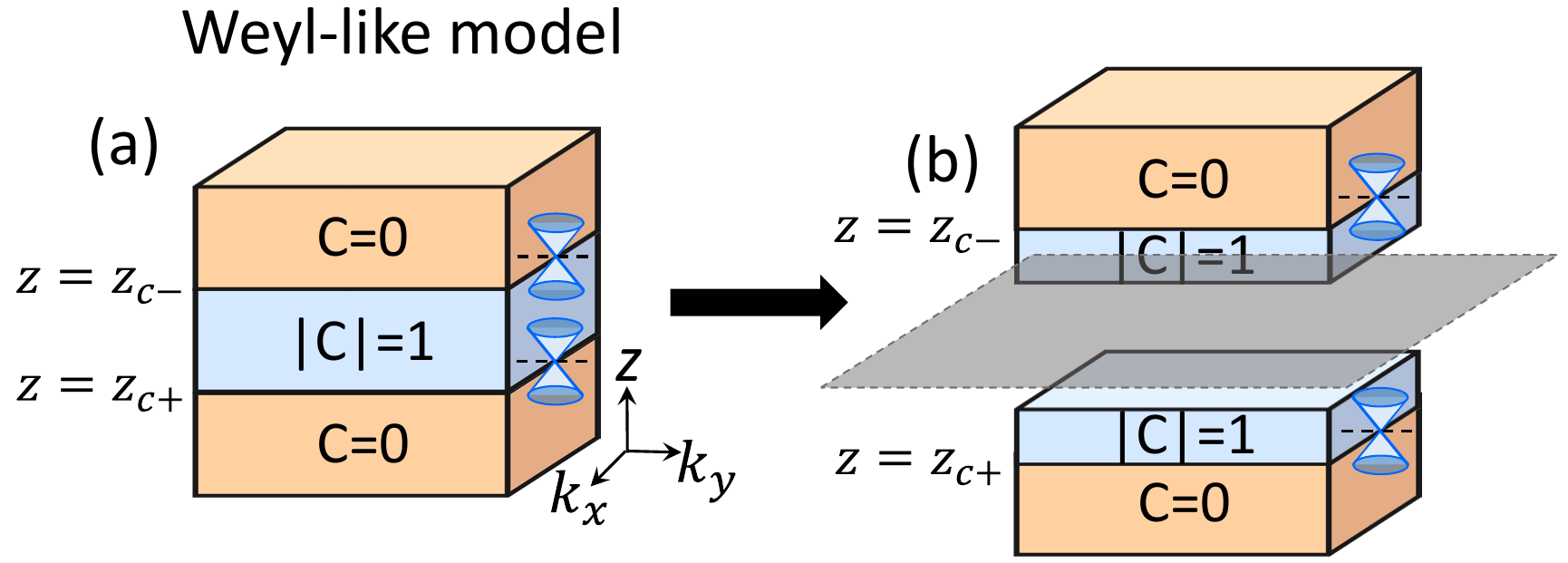}
\caption{(a) Weyl-like nodes at $z\!=\!z_{c-}$ and $z\!=\!z_{c+}$, marking topological transitions. (b) The two Weyl-like nodes are spatially separated by the bulk gap between $z\!=\!z_{c-}$ and $z\!=\!z_{c+}$. Each half of the system contains a single Weyl-like node. This differs from conventional Weyl points, which typically appear in pairs in momentum space $(k_x, k_y, k_z)$.}
\label{fig_halfcut}
\end{figure}  

\subsection{Comparison between the two models}\label{2.3}

As shown in Fig.~\ref{fig_dual}, the two models, the decoupled Weyl-like model (left column) and the Weyl semimetal model (right column), can be viewed as duals of each other.
The Weyl-like model in Eq.~\eqref{eq:H_z} provides a generic framework for describing Weyl semimetals, including key features such as band touching, Chern number, and Fermi arc, all of which are captured by the Weyl semimetal model in Eq.~\eqref{eq:H_Weyl}. 

The Weyl-like nodes do not obey the no-go theorem that constrains ordinary Weyl points. According to the no-go theorem~\cite{nielsen1981absence1,nielsen1981absence2}, Weyl points always appear in pairs. However, Fig.~\ref{fig_halfcut} illustrates that the two Weyl-like nodes are spatially separated by a bulk gap between $z\!=\!z_{c-}$ and $z\!=\!z_{c+}$, with each half of the system hosting a single Weyl-like node. This contrasts with conventional Weyl points, which, due to the no-go theorem~\cite{nielsen1981absence1,nielsen1981absence2}, typically appear in pairs in momentum space with opposite chiralities. In the next section, we turn our attention to the effects of interlayer coupling on the Weyl-like model Hamiltonian in Eq.~\eqref{eq:H_z}.

By comparing Eqs.~\eqref{eq:conductivity3D_last} and \eqref{eq:conductivity3D_Weyl_last}, we see that the 3D Hall conductivity in the Weyl-like model explicitly depends on the system size $L_z$, while that of the Weyl semimetal model is independent of system size. As such, the finite-size dependence of the 3D Hall conductivity cannot be captured by the Weyl semimetal model. This unique feature provides a clear signature of mixed-space topology in the Weyl-like model.

\subsection{Local Berry curvature}\label{2.4}

In this subsection, we calculate the local Berry curvature of the Weyl-like model in Eq.~\eqref{eq:H_z}.

To present the vector plot of the Berry curvature in the mixed space $(k_x,k_y,z)$, we first define the Berry curvature vector as~\cite{xiao2010berry,li2016negative,lu2017quantum}
\begin{eqnarray}
{\bf F}^{mn}({\bf k}_{||},z)\!=\!\nabla\!\times\!{\bf A}^{mn}({\bf k}_{||},z),\label{eq:Berry3D}
\end{eqnarray} where ${\bf A}^{mn}({\bf k}_{||},z)\!=\!i\langle\psi_{m}|\nabla\psi_{n}\rangle\!=\!({\cal A}_{k_x}^{mn},{\cal A}_{k_y}^{mn},{\cal A}_{z}^{mn})$ is the Berry connection vector in the mixed space $(k_x,k_y,z)$.

\begin{figure}
\centering
\includegraphics[width=0.65\columnwidth]{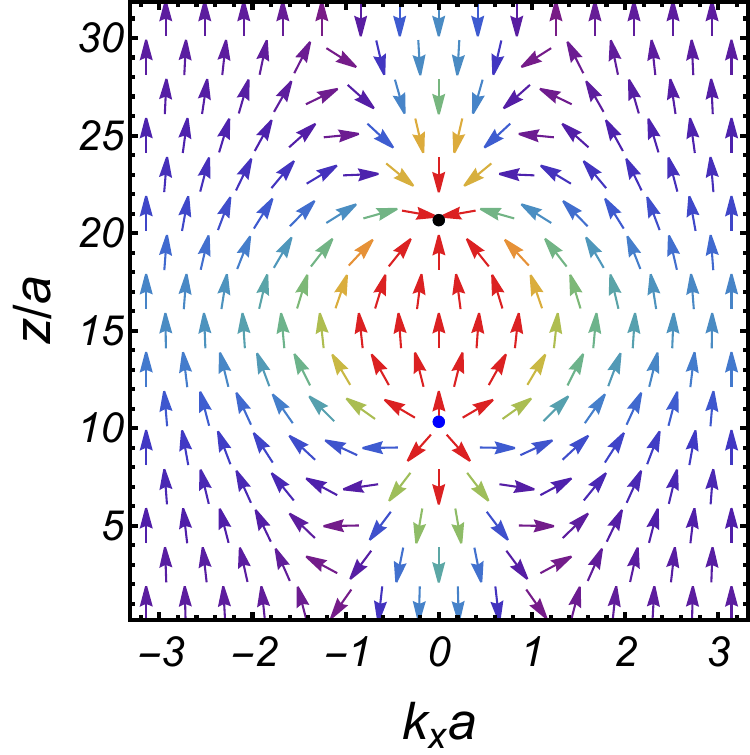}
\caption{Vector plot of the Berry curvature ${\bf F}({\bf k}_{||},z)$ [Eq.~\eqref{eq:Berry3D}] (summed over all occupied states $m,n$) in the mixed space $(k_x, z)$ with $k_y\!=\!0$ fixed for simplicity. A pair of Berry monopoles~\cite{li2016negative,lu2017quantum} is located at the two Weyl-like nodes. Arrows indicate the flux of the Berry curvature, flowing from one monopole [$(k_x\!=\!0, z_{c-}\!=\!10\frac{1}{3}a)$, blue point] to the other [$(k_x\!=\!0, z_{c+}\!=\!20\frac{2}{3}a)$, black point], revealing the nontrivial topological nature of the Weyl-like model in Eq.~\eqref{eq:H_z}. The parameters used are $m_{z}\!=\!0.5$ eV, $m_{0}\!=\!1$ eV, $V_{z}\!=\!1$ eV, $t_{x}\!=\!t_{y}\!=\!1$ eV, and $L_{z}\!=\!\lambda_{z}\!=\!31a$.}
\label{fig_3D_Berry_monopole}
\end{figure}

Figure \ref{fig_3D_Berry_monopole} shows the vector plot of the Berry curvature ${\bf F}({\bf k}_{||},z)$ in the mixed space $(k_x, z)$, with $k_y\!=\!0$ fixed for simplicity. A pair of Berry monopoles~\cite{li2016negative,lu2017quantum} is located at the two Weyl-like nodes. The arrows indicate the flux of the Berry curvature, which flows from one monopole [$(k_x\!=\!0, z_{c-}\!=\!10\frac{1}{3}a)$, blue point] to the other [$(k_x\!=\!0, z_{c+}\!=\!20\frac{2}{3}a)$, black point], reflecting the nontrivial topological nature of the Weyl-like model in Eq.~\eqref{eq:H_z}.

The chirality of a Weyl-like node can be obtained by integrating the Berry curvature over a Fermi surface enclosing the node~\cite{li2016negative,lu2017quantum}. This procedure yields opposite topological charges for the two Weyl-like points.

\begin{figure*}[h!tpb]
\centering
\includegraphics[width=0.8\textwidth]{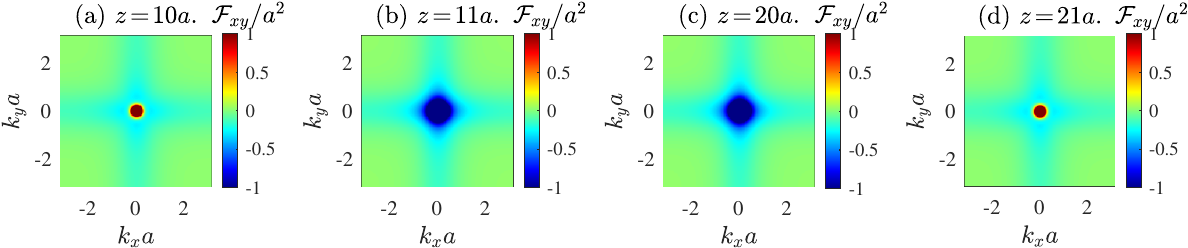}
\caption{Momentum distribution of the layer Berry curvature ${\cal F}_{xy}({\bf k}_{||},z)$ [Eq.~\eqref{eq:Berry_z}] (summed over all occupied states $m,n$) in the $(k_{x}, k_{y})$ plane for different layers: $j_z\!=\!z/a\!=\!10$, $11$, $20$, and $21$ (from left to right). (a) $z\!=\!10a$, (b) $z\!=\!11a$, (c) $z\!=\!20a$, and (d) $z\!=\!21a$. Here, $a$ is the lattice constant, and ${\cal F}_{xy}$ denotes the layer Berry curvature. The two critical planes are located at $z_{c-}\!=\!10\frac{1}{3}a$ and $z_{c+}\!=\!20\frac{2}{3}a$, given by $z_{c\pm}\!=\!\frac{\lambda_{z}}{2\pi}\left[\pi\pm\arccos\left(\frac{m_{z}}{V_{z}}\right)\right]$ [see Eq.~\eqref{eq:z_c}]. The parameters used are $m_{z}\!=\!0.5$ eV, $m_{0}\!=\!1$ eV, $V_{z}\!=\!1$ eV, $t_{x}\!=\!t_{y}\!=\!1$ eV, and $L_{z}\!=\!\lambda_{z}\!=\!31a$.}
\label{fig_Berry}
\end{figure*}

Furthermore, we investigate the momentum distribution of the layer Berry curvature ${\cal F}_{xy}({\bf k}_{||},z)$ near the two Weyl-like points $z_{c\pm}$.

Figure~\ref{fig_Berry} shows the momentum distribution of the layer Berry curvature ${\cal F}_{xy}({\bf k}_{||},z)$ in the $(k_x, k_y)$ plane for different layers: (a) $z\!=\!10a$, (b) $z\!=\!11a$, (c) $z\!=\!20a$, and (d) $z\!=\!21a$. Here, $a$ is the lattice constant. The two critical planes are located at $z_{c-}\!=\!10\frac{1}{3}a$ and $z_{c+}\!=\!20\frac{2}{3}a$, given by $z_{c\pm}\!=\!\frac{\lambda_z}{2\pi} \left[\pi \pm \arccos\left(\frac{m_z}{V_z}\right)\right]$ [see Eq.~\eqref{eq:z_c}], with parameters $m_z\!=\!0.5$ eV, $m_{0}\!=\!1$ eV, $V_z\!=\!1$ eV, $t_x\!=\!t_y\!=\!1$ eV, and $L_z\!=\!\lambda_z\!=\!31a$. Thus, $z\!=\!10a$ and $z\!=\!11a$ are close to the lower critical plane $z_{c-}$, while $z\!=\!20a$ and $z\!=\!21a$ are near the upper critical plane $z_{c+}$.

As shown in Fig.~\ref{fig_Berry}, the sign of the layer Berry curvature changes as the layer index increases from $z\!=\!10a$ [a topologically trivial region, see Fig.~\ref{fig_dual}(c1)] to $z\!=\!11a$ [a topologically nontrivial region, also shown in Fig.~\ref{fig_dual}(c1)]. This sign change corresponds to the band inversion occurring between these two layers, as illustrated in Fig.~\ref{fig_dual}(d1). Specifically, the energy gap present at $z\!=\!10a$ closes at the lower critical plane $z\!=\!z_{c-}$ and reopens at $z\!=\!11a$, signaling a topological phase transition.

Furthermore, the momentum distribution of the layer Berry curvature at $z\!=\!11a$ is more broadly concentrated around the center of the BZ ($k_x\!=\!k_y\!=\!0$) than that at $z\!=\!10a$. These observations are consistent with the Chern number being 1 in the region between the two critical planes ($z_{c-}\!<\!z\!<\!z_{c+}$) and 0 outside this region ($z\!<\!z_{c-}$ and $z\!>\!z_{c+}$), as indicated in Fig.~\ref{fig_dual}(c1).

A similar analysis applies to the layer Berry curvature across the upper critical plane $z\!=\!z_{c+}$.

\section{Interlayer coupling effects}\label{3}

We now incorporate the effects of interlayer coupling into our analysis. The energy bands are still obtained by calculating the local density of states~\cite{chen2021field,wan2011topological,yang2011quantum}, while the quantum transport properties are captured by evaluating the layer Hall conductance~\cite{fukui2005chern,sun2020analytical,otrokov2019unique,gao2021layer,chen2022layer,mong2010antiferromagnetic,essin2009magnetoelectric,wang2015quantized,varnava2018surfaces,chen2023half,fu2021bulk,price2015four,zhang2019entangled} in the presence of interlayer coupling. Additionally, we examine how interlayer coupling influences the total 3D Hall conductivity.

The interlayer hopping term is given by
\begin{equation}
\Delta H\!=\!\sum_{j_{z}}\eta_{z}\sigma_{z}\left(C_{j_{z}+1}^{\dag}C_{j_{z}}^{}\!+\!C_{j_{z}}^{\dag}C_{j_{z}+1}^{}\right),\label{eq:Delta_H}
\end{equation} where $C_{j_{z}}^{\dag}$ ($C_{j_{z}}^{}$) is the electron creation (annihilation) operator on the $j_{z}$th layer, and $\eta_{z}\sigma_{z}$ represents the nearest-neighbor interlayer hopping matrix.
Combining Eq.~\eqref{eq:H_z} with the interlayer hopping term, the full Hamiltonian incorporating interlayer coupling becomes
\begin{equation}
H_{\rm int}\!=\!\sum_{j_{z}}\left[h_{\text{2D}}({\bf k}_{||})\!+\!V(z)\sigma_{z}\right]C_{j_{z}}^{\dag}C_{j_{z}}^{}\!+\!\Delta H. \label{eq:H_int}
\end{equation} The matrix representation of the Hamiltonian in Eq.~\eqref{eq:H_int} under OBCs along the $z$ direction is provided in Section SV in the Supplemental Material~\cite{SuppMat}. 

\subsection{Local density of states}\label{3.1}

\begin{figure}[h!tpb]
\centering
\includegraphics[width=\columnwidth]{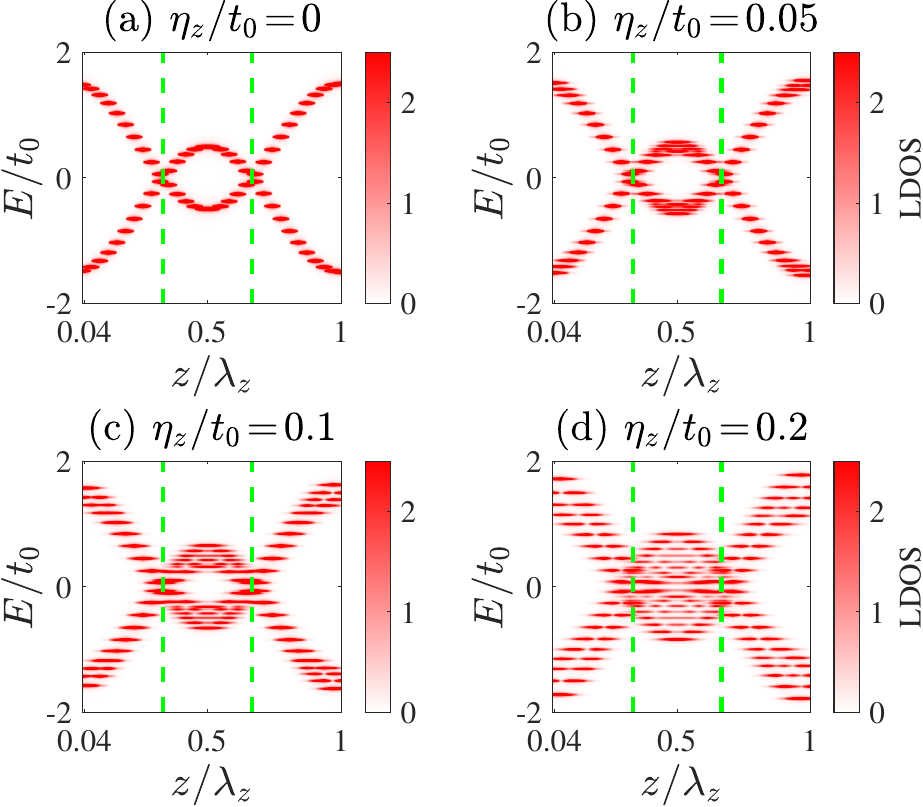}
\caption{Local density of states (LDOS) [Eq.~\eqref{eq:LDOS}] for different interlayer hopping strengths $\eta_{z}$. The LDOS is indicated by the red color, while the two green vertical dashed lines mark the positions of the Weyl-like points. (a) $\eta_{z}/t_{0}\!=\!0$, (b) $\eta_{z}/t_{0}\!=\!0.05$, (c) $\eta_{z}/t_{0}\!=\!0.1$, and (d) $\eta_{z}/t_{0}\!=\!0.2$. The other parameters are $k_x\!=\!k_y\!=\!0$, $k_{B}T=0.01$ eV, $V_z\!=\!1$ eV, $m_z\!=\!0.5$ eV, $m_0\!=\!1$ eV, $t_x\!=\!t_y\!=\!t_{0}\!=\!1$ eV, and $L_z\!=\!\lambda_z\!=\!31a$, with $a\!=\!1$ nm.}
\label{fig_LDOS}
\end{figure}

The local density of states is defined as~\cite{zhang2023emergent}
\begin{align}
\rho(z,E,T)\!=\!\sum_{i}\rho^{}_{i}(z)g(E,E_{i},T), \label{eq:LDOS}
\end{align} where the spatial density distribution $\rho_{i}^{}(z)$ of the $i$th eigenstates $|\psi_{i}(z)\rangle$ is given by
\begin{align}
\rho_{i}^{}(z)\!=\!\langle\psi_{i}^{}(z)|\psi_{i}^{}(z)\rangle,
\end{align} and $g(E,E_{n},T)$ is a Lorentzian broadening function defined as
\begin{align}
g(E,E_{n},T)\!=\!\lim_{k_{B}T\to0^{+}}\frac{k_{B}T}{\pi[(E\!-\!E_{n})^{2}\!+\!(k_{B}T)^2]}.
\end{align} 
Here, $E$ is the probe energy, $k_{B}$ is the Boltzmann constant, and $T$ is the temperature.

Furthermore, we numerically calculate the local density of states for different interlayer hopping strengths, as shown in Fig.~\ref{fig_LDOS}.
In the decoupled case with $\eta_{z}\!=\!0$, Fig.~\ref{fig_LDOS}(a) shows that the local density of states exhibits clear crossings at the two Weyl-like points. The corresponding probability distributions around these points are symmetric, with equal weight at each Weyl-like point.

For weak interlayer coupling $\eta_{z}/t_{0}\!=\!0.05$, Fig.~\ref{fig_LDOS}(b) reveals that the local density of states still exhibits sharp crossings at the same Weyl-like points, like the decoupled case in Fig.~\ref{fig_LDOS}(a). For moderate coupling, $\eta_{z}/t_{0}\!=\!0.1$, Fig.~\ref{fig_LDOS}(c) shows that the local density of states continues to display distinct crossings at the Weyl-like points, indicating that the features observed in the decoupled case persist.
These results suggest that the Weyl-like points are robust against weak to moderate interlayer coupling.
However, as the interlayer hopping strength $\eta_{z}$ increases further, Fig.~\ref{fig_LDOS}(d) shows that the local density of states becomes increasingly dispersive compared with the decoupled scenario. This behavior indicates that strong interlayer coupling disrupts the localized character of the Weyl-like points, ultimately leading to their destruction.

\subsection{Layer Hall conductance}\label{3.2}

\begin{figure}[h!tpb]
\centering
\includegraphics[width=\columnwidth]{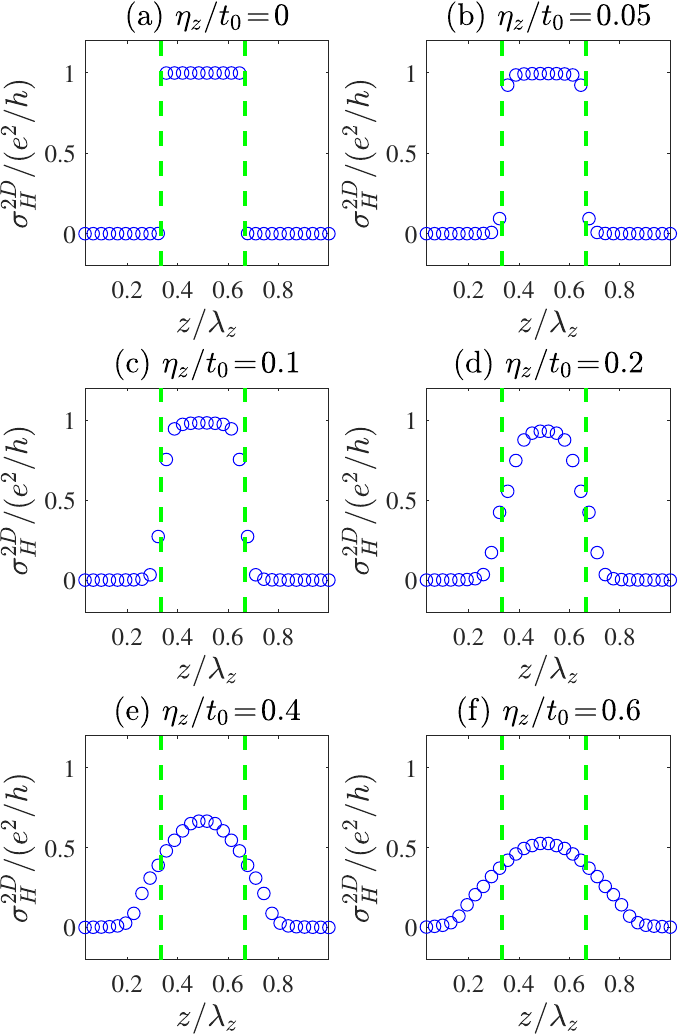}
\caption{Layer Hall conductances for different interlayer hopping strengths $\eta_{z}$. The two green vertical dashed lines indicate the positions of the Weyl-like points. (a) $\eta_{z}/t_{0}\!=\!0$, (b) $\eta_{z}/t_{0}\!=\!0.05$, (c) $\eta_{z}/t_{0}\!=\!0.1$, (d) $\eta_{z}/t_{0}\!=\!0.2$, (e) $\eta_{z}/t_{0}\!=\!0.4$, and (f) $\eta_{z}/t_{0}\!=\!0.6$. All other parameters are the same as those used in Fig.~\ref{fig_LDOS}.}
\label{fig_layer_Chern}
\end{figure}

To calculate the layer Hall conductance, we substitute the Berry connection from Eq.~\eqref{eq:Berry_connection_z} into Eqs.~\eqref{eq:Berry_z} and \eqref{eq:Chern_z}. The resulting layer Hall conductances for different interlayer hopping strengths $\eta_{z}$ are shown in Fig.~\ref{fig_layer_Chern}.

Comparing the decoupled case $\eta_{z}\!=\!0$ [Fig.~\ref{fig_layer_Chern}(a)] with the weakly coupled case $\eta_{z}/t_{0}\!=\!0.05$ [Fig.~\ref{fig_layer_Chern}(b)] and the moderately coupled case $\eta_{z}/t_{0}\!=\!0.1$ [Fig.~\ref{fig_layer_Chern}(c)], we observe that the layer Hall conductance is only slightly modified near the two Weyl-like points. This result demonstrates that the Weyl-like points remain robust under weak and moderate interlayer couplings.

However, as the interlayer hopping strength $\eta_{z}$ increases further [Figs.~\ref{fig_layer_Chern}(d)--\ref{fig_layer_Chern}(f)], the layer Hall conductance becomes increasingly smooth and continuous in the vicinity of the Weyl-like points, deviating from the sharp features observed in the decoupled case. Physically, this behavior indicates that strong interlayer coupling destroys the Weyl-like points and significantly alters the bulk topology of the system.

\subsection{Total 3D Hall conductivity}\label{3.3}

\begin{figure}[h!tpb]
\centering
\includegraphics[width=\columnwidth]{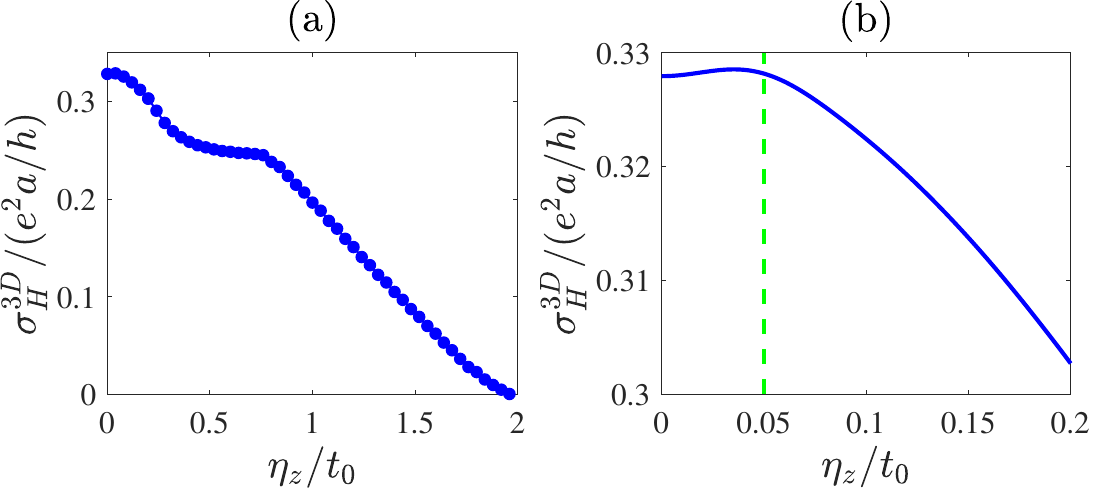}
\caption{Total 3D Hall conductivity [Eq.~\eqref{eq:conductivity3D_Weyl_like}] as a function of the interlayer hopping strength $\eta_{z}$. (a) Full range: $\eta_{z}/t_{0} \in [0, 2]$. (b) Enlarged view of the moderate-coupling regime: $\eta_{z}/t_{0} \in [0, 0.2]$. Here, $L_z\!=\!\lambda_z\!=\!61a$, and the other parameters are the same as those used in Fig.~\ref{fig_LDOS}.}
\label{fig_Hall3D}
\end{figure}

To investigate the effects of interlayer coupling on the total 3D Hall conductivity, we compute it as a function of the interlayer hopping strength $\eta_{z}$, as shown in Fig.~\ref{fig_Hall3D}(a), based on Eq.~\eqref{eq:conductivity3D_Weyl_like}. The results indicate that, for $\eta_{z}/t_{0}>0.05$, the total 3D Hall conductivity decreases monotonically as $\eta_{z}$ increases, eventually approaching zero when $\eta_{z} \sim 2t_{0}$. This behavior indicates that strong interlayer coupling can effectively suppress the total 3D Hall conductivity.

A closer look at the range $\eta_{z}/t_{0} \in [0, 0.2]$ [Fig.~\ref{fig_Hall3D}(b)] reveals an approximate plateau within $\eta_{z}/t_{0} \in [0, 0.05]$. This suggests that the total 3D Hall conductivity is robust against weak interlayer coupling.

\section{Disorder}\label{4}

To examine the effect of a generic disorder potential on the Weyl-like points and the associated topological invariant, we introduce an Anderson-type disorder~\cite{li2009topological,groth2009theory,stutzer2018photonic,meier2018observation,li2020topological,qin2023universal,chen2024tunable,chen2017disorder,chen2017topological,chen2018floquet2,chen2018phase,yi2024disorder,chen2025real}
\begin{eqnarray}\label{eq:H_disorder}
H_{d}\!=\!\sum_{j_x,j_y,j_z}W(j_x,j_y,j_z)\sigma_{0}^{}C_{j_x,j_y,j_z}^{\dagger}C_{j_x,j_y,j_z}^{},
\end{eqnarray}
where $C_{j_x,j_y,j_z}^{\dag}$ ($C_{j_x,j_y,j_z}^{}$) denotes the electron creation (annihilation) operator at site $(j_x,j_y,j_z)$, the summation runs over all lattice sites, and $\sigma_{0}^{}$ is a $2\times2$ identity matrix.
The onsite disorder potential $W(j_x,j_y,j_z)$ is a random variable uniformly distributed in the energy interval $[-U,U]$, where $U$ measures the disorder strength. 

Based on the model Hamiltonian in Eq.~\eqref{eq:H_z}, we emphasize that the system is decoupled along the 
$z$ direction and can therefore be decomposed into multiple 2D models for computation.

To incorporate the Anderson-type disorder in Eq.~\eqref{eq:H_disorder} into our model Hamiltonian, we rewrite the Hamiltonian in real space as follows.
By performing 2D Fourier transforms, we obtain the real-space disordered Weyl-like model Hamiltonian with PBCs in the $x$ and $y$ directions. In the basis
\begin{eqnarray}
&&(C_{1,1,j_z}, C_{1,2,j_z},\cdots, C_{1,L_y,j_z}, C_{2,1,j_z}, C_{2,2,j_z},\cdots, C_{2,L_y,j_z},\nonumber\\
&&\cdots, C_{L_x,L_y,j_z})^{T},\nonumber
\end{eqnarray} the Hamiltonian reads
\begin{eqnarray}\label{eq:Hxyz_disorder}
{\cal H}_{yx}\!=\!\begin{pmatrix}
h_{1} & T_{1} & 0 & \cdots & T_{1}^{\dagger} \\
T_{1}^{\dagger} & h_{1} & T_{1} & \cdots & 0 \\
0 & T_{1}^{\dagger} & h_{1} & \ddots & \vdots \\
\vdots& \ddots & \ddots & \ddots & T_{1} \\
T_{1} & \cdots & 0 & T_{1}^{\dagger} & h_{1}
\end{pmatrix}_{(2L_yL_x)\times (2L_yL_x)},
\end{eqnarray} where
\begin{eqnarray}
h_{1}\!=\!\begin{pmatrix}
M(z) & T_{y} & 0 & \cdots & T_{y}^{\dagger} \\
T_{y}^{\dagger} & M(z) & T_{y} & \cdots & 0 \\
0 & T_{y}^{\dagger} & M(z) & \ddots & \vdots \\
\vdots& \ddots & \ddots & \ddots & T_{y} \\
T_{y} & \cdots & 0 & T_{y}^{\dagger} & M(z)
\end{pmatrix}_{2L_y\times 2L_y},
\end{eqnarray}
\begin{eqnarray}
T_{1}\!=\!\begin{pmatrix}
T_{x} & 0 & 0 & \cdots & 0 \\
0 & T_{x} & 0 & \cdots & 0 \\
0 & 0 & T_{x} & \ddots & \vdots \\
\vdots& \ddots & \ddots & \ddots & 0 \\
0 & \cdots & 0 & 0 & T_{x} 
\end{pmatrix}_{2L_y\times 2L_y}.
\end{eqnarray} The terms $M(z)$, $T_x$, and $T_y$ are given by
\begin{eqnarray}
&&M(z)\!=\![m_{z}\!+\!V(z)\!+\!2m_{0}]\sigma_{z} \!+\! W(j_x,j_y,j_z)\sigma_{0}^{},\\
&&T_{x}\!=\!-\frac{m_{0}}{2}\sigma_{z} - i\frac{t_{x}}{2}\sigma_{x},~~
T_{x}^{\dagger}\!=\!-\frac{m_{0}}{2}\sigma_{z} + i\frac{t_{x}}{2}\sigma_{x},\\
&&T_{y}\!=\!-\frac{m_{0}}{2}\sigma_{z} - i\frac{t_{y}}{2}\sigma_{y},~~
T_{y}^{\dagger}\!=\!-\frac{m_{0}}{2}\sigma_{z} + i\frac{t_{y}}{2}\sigma_{y},
\end{eqnarray} where $z\!=\!j_{z}a$.

The detailed derivation of the real-space disordered Weyl-like model Hamiltonian in Eq.~\eqref{eq:Hxyz_disorder} with PBCs in the $x$- and $y$-directions is provided in Section SVI in the Supplemental Material~\cite{SuppMat}

\subsection{Local Chern marker}\label{4.1}

\begin{figure}
\centering
\includegraphics[width=\columnwidth]{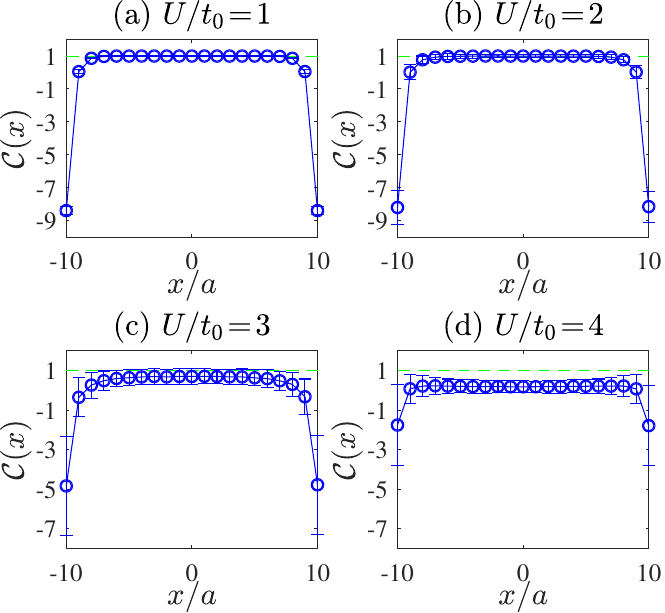}
\caption{Local Chern marker [Eq.~\eqref{eq:local_Chern_marker}] of the real-space disordered Weyl-like model Hamiltonian in Eq.~\eqref{eq:Hxyz_disorder}, evaluated at fixed $y\!=\!0$ and $z\!=\!16a$ (the center position along the $z$ direction with $L_{z}\!=\!\lambda_{z}\!=\!31a$), for different disorder strengths $U$. (a) $U/t_{0}\!=\!1$, (b) $U/t_{0}\!=\!2$, (c) $U/t_{0}\!=\!3$, and (d) $U/t_{0}\!=\!4$. The data with error bars represent the standard deviation of the local Chern marker calculated over 1000 samples. The parameters used are $m_{z}\!=\!0.5$ eV, $m_{0}\!=\!1$ eV, $V_{z}\!=\!1$ eV, $t_{x}\!=\!t_{y}\!=\!1$ eV, $L_{x}\!=\!L_{y}\!=\!21a$, and $a\!=\!1$ nm.}
\label{fig_WLP_local_Chern_marker_x}
\end{figure}

To investigate the topological invariant in disordered systems, we employ the local Chern marker, defined as~\cite{bianco2011mapping,caio2019topological,li2021two}
\begin{eqnarray}
{\cal C}(m)\!=\!-\frac{4\pi}{A_{c}}{\rm Im}\sum_{s=A,B}\langle{\bf r}_{m,s}|\hat{P}\hat{x}\hat{Q}\hat{y}\hat{P}|{\bf r}_{m,s}\rangle,\label{eq:local_Chern_marker}
\end{eqnarray}
where $A_{c}$ is the area of a real-space unit cell, and $|{\bf r}_{m,s}\rangle$ denotes the sublattice state $s$ ($s\!=\!A,B$) in the $m$th unit cell, labeled by $m\!=\!(j_x,j_y)$. The operator $\hat{P}\!=\!\sum_{n}|\psi_{n}\rangle\langle\psi_{n}|$ is the projector onto the chosen quasienergy band in real space, while $\hat{Q}\!=\!\hat{I}\!-\!\hat{P}$ is its complement, with $\hat{I}$ being the identity operator. The operators $\hat{x}$ and $\hat{y}$ represent the position operators along the $x$ and $y$ directions, respectively.

Furthermore, we plot the local Chern marker for the real-space disordered Weyl-like model Hamiltonian in Eq.~\eqref{eq:Hxyz_disorder}, evaluated at fixed $y\!=\!0$ and $z\!=\!16a$ (the center position along the $z$ direction with $L_{z}\!=\!\lambda_{z}\!=\!31a$), for various disorder strengths $U$, as shown in Fig.~\ref{fig_WLP_local_Chern_marker_x}.
By comparing Figs.~\ref{fig_WLP_local_Chern_marker_x}(a) and \ref{fig_WLP_local_Chern_marker_x}(b), we observe that the local Chern numbers ${\cal C}(x)$ in the bulk of the system fluctuate around a macroscopic average of ${\cal C}(x)\!=\!1$ for disordered cases with weak to moderate disorder (up to $U/t_{0}\!\approx\!2$).
Particularly, the panels demonstrate that the value of the local Chern marker remains stable at the center of the $(x, y)$ plane, i.e., at $(x\!=\!0, y\!=\!0)$.
Therefore, in the following discussions, we focus on the local Chern marker at this central point.

\subsection{Layer Hall conductance with disorder}\label{4.2}

\begin{figure}[h!tpb]
\centering
\includegraphics[width=\columnwidth]{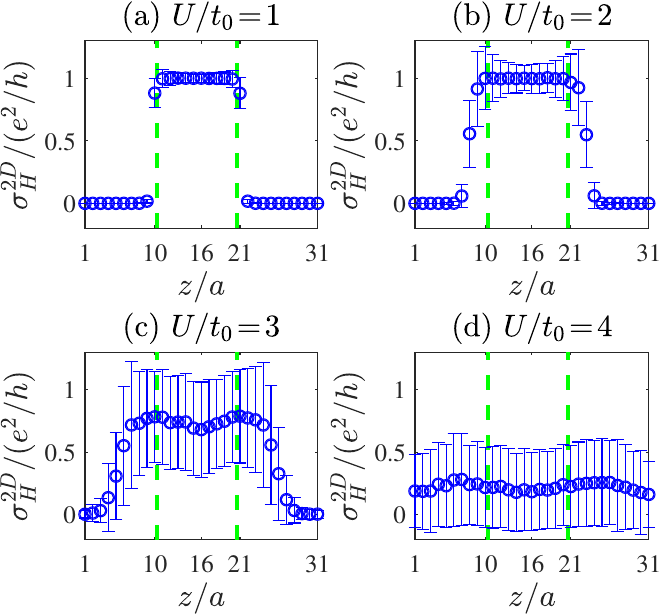}
\caption{Layer Hall conductances of the real-space disordered Weyl-like model Hamiltonian in Eq.~\eqref{eq:Hxyz_disorder} for different disorder strengths $U$. The two green vertical dashed lines indicate the positions of the Weyl-like points. The data with error bars represent the standard deviation of the Hall conductance calculated over 1000 samples. (a) $U/t_{0}\!=\!1$, (b) $U/t_{0}\!=\!2$, (c) $U/t_{0}\!=\!3$, and (d) $U/t_{0}\!=\!4$. Here, we use the local Chern marker evaluated at $x\!=\!y\!=\!0$. The system length along the $z$ direction is $L_z\!=\!\lambda_z\!=\!31a$, and all other parameters are the same as those in Fig.~\ref{fig_WLP_local_Chern_marker_x}.}
\label{fig_WLP_layer_local_Chern_marker}
\end{figure}

Based on the local Chern marker in Eq.~\eqref{eq:local_Chern_marker}, we plot the layer Hall conductances of the real-space disordered Weyl-like model Hamiltonian in Eq.~\eqref{eq:Hxyz_disorder} for various disorder strengths, as shown in Fig.~\ref{fig_WLP_layer_local_Chern_marker}.

Figure~\ref{fig_WLP_layer_local_Chern_marker} reveals that, although the fluctuations increase with growing disorder strength, the two Weyl-like points remain robust under weak to moderate disorder (up to $U/t_{0}\!\approx\!2$). Moreover, the Hall conductance exhibits pronounced fluctuations near the topological phase transition points, i.e., the Weyl-like points $z_{c\pm}$, even at fixed weak or moderate disorder strengths ($U/t_{0}\!\leqslant\!2$).

\subsection{3D Hall conductivity with disorder}\label{4.3}

\begin{figure}[h!tpb]
\centering
\includegraphics[width=0.8\columnwidth]{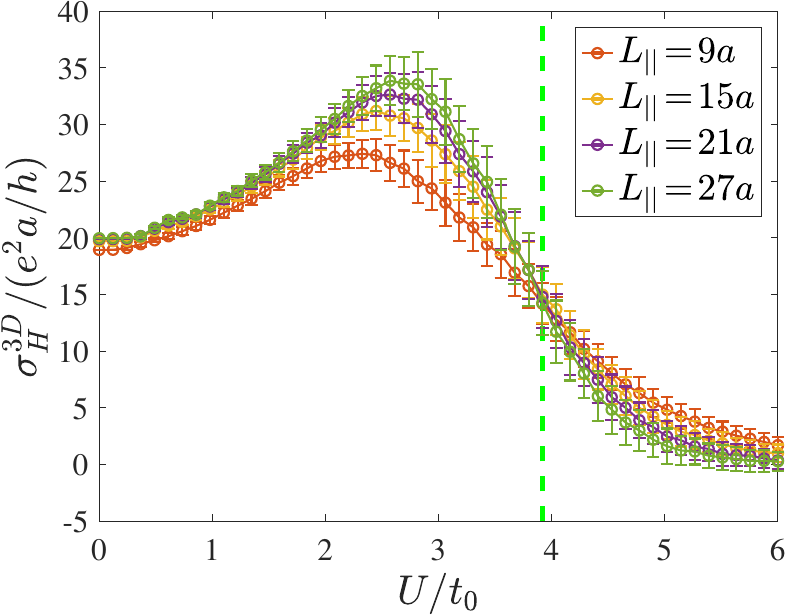}
\caption{3D Hall conductivity [Eq.~\eqref{eq:conductivity3D_Weyl_like}] as a function of the disorder strength $U$ for different system sizes $L_{||}\!=\!9a$, $15a$, $21a$, and $27a$, where $L_{||}\!=\!L_{x}\!=\!L_{y}$. The data with error bars represent the standard deviation of the 3D Hall conductivity calculated over 1000 samples. The system length along the $z$-direction is $L_z\!=\!\lambda_z\!=\!61a$, and all other parameters are the same as those used in Fig.~\ref{fig_WLP_layer_local_Chern_marker}.}
\label{fig_Hall3D_disorder}
\end{figure}

To investigate the critical disorder threshold at which the topological phase becomes unstable, we focus on the 3D Hall conductivity as a function of the disorder strength $U$, for different system sizes $L_{||}\!=\!L_{x}\!=\!L_{y}$ in the $x$ and $y$ directions, as shown in Fig.~\ref{fig_Hall3D_disorder}.

Interestingly, an inverse point is observed at $U_{\rm inv}/t_{0}\!\approx\!3.9$ (the green vertical dashed line) across different system sizes $L_{||}$, as shown in Fig.~\ref{fig_Hall3D_disorder}. For $U\!<\!U_{\rm inv}$, the 3D Hall conductivity increases with system size $L_{||}$, whereas for $U\!>\!U_{\rm inv}$, it decreases as $L_{||}$ grows. In the latter case, as $L_{||}$ becomes large, the 3D Hall conductivity approaches zero, indicating that beyond this disorder strength $U_{\rm inv}$, the topological phase tends to become unstable.

\subsection{Born approximation}\label{4.4}

\begin{figure}[h!tpb]
\centering
\includegraphics[width=0.8\columnwidth]{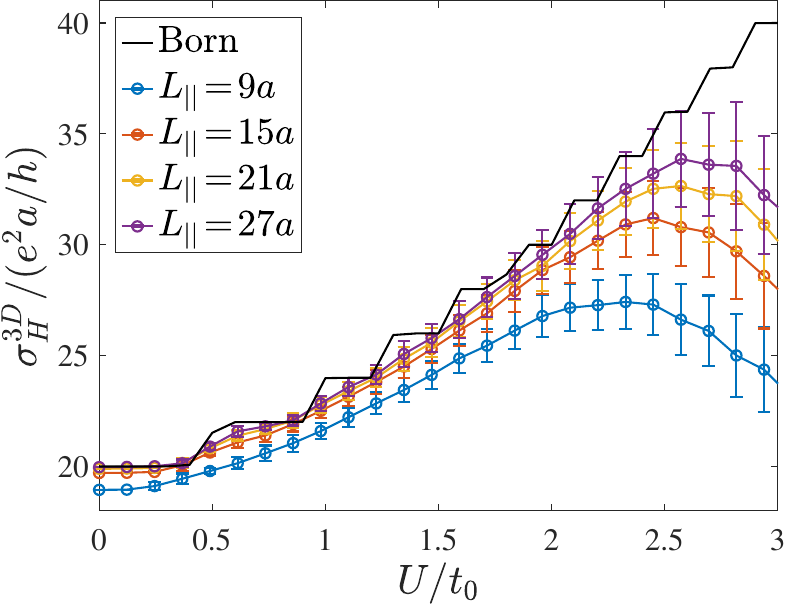}
\caption{3D Hall conductivity [Eq.~\eqref{eq:conductivity3D_Weyl_like}] as a function of the disorder strength $U$. The black solid curve denotes the result obtained using the Born approximation [Eq.~\eqref{eq:Sigma1}]. The data with error bars represent the standard deviation of the 3D Hall conductivity calculated over 1000 samples. All other parameters are the same as those used in Fig.~\ref{fig_Hall3D_disorder}.}
\label{fig_Hall3D_disorder_Born}
\end{figure}

To support the physical interpretation of the numerical simulations for the 3D Hall conductivity, as shown in Fig.~\ref{fig_Hall3D_disorder}, we analyze the present model using an effective medium theory based on the Born approximation, which neglects higher-order scattering processes~\cite{groth2009theory,chen2017disorder,chen2017topological,chen2018floquet2,chen2018phase,yi2024disorder,chen2025real}. Within the self-consistent Born approximation, the self-energy $\Sigma$ for a finite disorder strength is determined by the following integral equation~\cite{ominato2014quantum,song2012dependence,hung2016disorder,guo2010topological,groth2009theory,chen2017disorder,chen2017topological,chen2018floquet2,chen2018phase,yi2024disorder,chen2025real}:
\begin{eqnarray}
&&\Sigma(z)\!=\!\frac{U^{2}}{3}\left(\frac{a}{2\pi}\right)^{d} \nonumber\\
&&\times\!\!\!\int_{\rm BZ}\!\!\left\{\!\sigma_{s}\!\left[ (E_{F}\!+\!i\delta)\mathbb{I}\!-\!{\cal H}({\bf k}_{||},z)\!-\!\Sigma(z) \right]^{-1}\!\sigma_{s}\!\right\}d^{2}{\bf k}_{||}, \label{eq:Sigma0}\nonumber\\
\end{eqnarray}
where $E_F$ is the Fermi energy, $\delta$ is a positive infinitesimal, i.e., $\delta\!\to\!0^{+}$, $d\!=\!2$ is the dimension of the momentum space, $\sigma_{s}\!=\!\sigma_{0/x}$ denotes the type of disorder~\cite{song2012dependence,hung2016disorder}---$\sigma_{s}\!=\!\sigma_{x}$ when the disorder is on the hopping terms and $\sigma_{s}\!=\!\sigma_{0}^{}$ when the disorder is on-site~\cite{hung2016disorder}---${\cal H}({\bf k}_{||},z)$ is the Weyl-like model Hamiltonian in Eq.~\eqref{eq:H_z}, and $\mathbb{I}$ is a $2\times2$ identity matrix. 
The coefficient $\frac{1}{3}$ arises from the variance $\langle U^{2}\rangle\!=\!U^{2}/3$ of a random variable uniformly distributed over the interval $[-U,U]$. The integration is performed over the first BZ. In the lowest-order Born approximation, we set $\Sigma(z)\!=\!0$ on the right-hand side of Eq.~\eqref{eq:Sigma0}. Therefore, for the on-site disorder [Eq.~\eqref{eq:H_disorder}] and $d\!=\!2$, we have 
\begin{eqnarray}
\Sigma(z)
\!\approx\!\frac{U^{2}}{3}\left(\frac{a}{2\pi}\right)^{2} 
\!\!\!\int_{\rm BZ}\!\left[ (E_{F}\!+\!i\delta)\mathbb{I}\!-\!{\cal H}({\bf k}_{||},z) \right]^{-1}d^{2}{\bf k}_{||}. \label{eq:Sigma1}\nonumber\\
\end{eqnarray}
The Hamiltonian matrix ${\cal H}({\bf k}_{||},z)\!=\!\sum_{i=x,y,z}h_{i}\sigma_{i}$ is $2\times2$, so that the integrand in Eq.~\eqref{eq:Sigma1} is written as~\cite{hung2016disorder}
\begin{eqnarray}
\left[ (E_{F}\!+\!i\delta)\mathbb{I}\!-\!{\cal H}({\bf k}_{||},z) \right]^{-1}\!=\!\frac{(E_{F}\!+\!i\delta)\mathbb{I}\!+\!{\cal H}({\bf k}_{||},z)}{(E_{F}\!+\!i\delta)^{2}\!-\!\sum_{i=x,y,z}h_{i}^{2}}, \nonumber\\
\end{eqnarray} which is more convenient for our numerical calculations.

By analyzing the renormalized Hamiltonian ${\cal H}({\bf k}_{||},z)\!+\!\Sigma(z)$, we find that the Born approximation results agree well with the numerical calculations in the weak to moderate disorder regime ($U/t_{0}\!\leqslant\!2$), as shown in Fig.~\ref{fig_Hall3D_disorder_Born}, where the black solid line denotes the prediction from the self-consistent Born approximation.

\section{Floquet Hamiltonian}\label{5}

To investigate the Floquet engineering of the two models discussed in Section~\ref{2}, we derive their effective Floquet Hamiltonians.

The optical field propagating along the $z$ direction in the material is given by ${\bf E}(t)\!=\!\partial{\bf A}(t)/\partial t\!=\!E_{0}(\cos(\omega t),\cos(\omega t \!+\! \varphi),0)$, where $E_{0}$ is the amplitude, $\omega$ is the angular frequency, and $\varphi$ represents the phase. For linear polarization, $\varphi\!=\!0$, while $\varphi\!=\!\mp\pi/2$ corresponds to left- or right-handed circular polarization. The vector potential is periodic with period $T\!=\!2\pi/\omega$, and it is given by ${\bf A}(t)\!=\!{\bf A}(t\!+\!T)\!=\!\omega^{-1}E_{0}(\sin(\omega t),\sin(\omega t \!+\! \varphi),0)$.
In the off-resonant regime, where the central Floquet band is well-separated from other replicas, the high-frequency expansion applies. We choose the optical frequency as $\hbar\omega\!=\!10$ eV, which is much larger than the  bandwidth of the material~\cite{zhan2024perspective,qin2023light,qin2022light,qin2022phase,qin2024light,dabiri2021light,dabiri2021engineering,dabiri2022floquet,askarpour2022light,pervishko2018impact}.

Under optical driving, the electron motion is modified by minimal coupling to the electromagnetic gauge field ${\bf A}(t)$, resulting in the photon-dressed effective Hamiltonian:
\begin{eqnarray}\label{eq:Ht}
{\cal H}({\bf k},t)\!=\!{\cal H}\left({\bf k} \!-\! \frac{e}{\hbar}{\bf A}(t)\right).
\end{eqnarray} 
Next, we derive the effective Floquet Hamiltonian~\cite{oka2009photovoltaic,calvo2015floquet,seshadri2022engineering,seshadri2019generating,seshadri2022floquet,lee2018floquet} \begin{eqnarray}
{\cal H}^{(F)}({\bf k})\!=\!\frac{i}{T}\ln\left\{\mathcal{T}\exp\left[-i\int^T_0 {\cal H}({\bf k},t)dt\right]\right\},
\end{eqnarray} where $\mathcal{T}$ is the time-ordering operator. In the high-frequency regime, a closed-form solution exists via the high-frequency expansion~\cite{magnus1954exponential,blanes2009magnus,lee2018floquet,bukov2015universal,eckardt2015high,chen2018floquet1,chen2018floquet2,du2022weyl,wang2022floquet} \begin{eqnarray}\label{eq:HF0} 
{\cal H}^{(F)}({\bf k}) \!=\! {\cal H}_{0} \!+\! \sum_{n=1}^{\infty}\frac{[{\cal H}_{-n}, {\cal H}_{n}]}{n\hbar\omega}\!+\! {\cal O}(\omega^{-2}),
\end{eqnarray} where ${\cal H}_{m-m'} \!=\! \frac{1}{T} \int_{0}^{T}{\cal H}({\bf k},t) \exp[i(m-m')\omega tdt]$, and $m$, $m'$ are integers. 

\subsection{Floquet Weyl-like model}\label{5.1}

From Eq.~\eqref{eq:HF0}, the Floquet Hamiltonian for the Weyl-like model can be expressed as
\begin{eqnarray}
{\cal H}^{(F)}({\bf k}_{||},z)\!=\!\sum_{i=x,y,z}h_{i}\sigma_{i},
\label{eq:H_F}
\end{eqnarray} where 
\begin{eqnarray}
h_{x}&\!=\!&{\cal J}_{0}(A_{0}a)t_{x}\sin(k_{x}a) \nonumber\\
&&\!-\!\!\!\!\!\!\!\sum_{n\in{\rm odd},n>0}\!\frac{4t_{x}m_{0}{\cal J}_{n}^{2}(A_{0}a)}{n\hbar\omega}\sin(n\varphi)\sin(k_{x}a)\cos(k_{y}a), \nonumber\\
\\
h_{y}&\!=\!&{\cal J}_{0}(A_{0}a)t_{y}\sin(k_{y}a) \nonumber\\
&&\!-\!\!\!\!\!\!\!\sum_{n\in{\rm odd},n>0}\!\frac{4t_{x}m_{0}{\cal J}_{n}^{2}(A_{0}a)}{n\hbar\omega}\sin(n\varphi)\cos(k_{x}a)\sin(k_{y}a), \nonumber\\
\\
h_{z}&\!=\!&m_{z}\!+\!V_{z}\cos(Q_{z}z)\!+\!2m_{0}[1-{\cal J}_{0}(A_{0}a)] \nonumber\\
&&\!+{\cal J}_{0}(A_{0}a)m_{0}\left[2\!-\!\cos(k_{x}a) \!-\! \cos(k_{y}a)\right] \nonumber\\
&&\!+\!\!\!\!\!\!\!\sum_{n\in{\rm odd},n>0}\!\frac{4t_{x}t_{y}{\cal J}_{n}^{2}(A_{0}a)}{n\hbar\omega}\sin(n\varphi)\cos(k_{x}a)\cos(k_{y}a),\nonumber\\
\end{eqnarray} where ${\cal J}_{n}(A_{0}a)$ is the $n$th Bessel function of the first kind~\cite{temme1996special}, and $A_{0}\!=\!eE_{0}/(\hbar\omega)$. The detailed derivations leading to Eq.~\eqref{eq:H_F} are provided in Subsection SVII A in the Supplemental Material~\cite{SuppMat}. 
For $\varphi\!=\!0$ (linearly polarized laser), the Weyl-like nodes are determined by solving the equation $m_{z}\!+\!V_{z}\cos(Q_{z}z)\!+\!2m_{0}[1\!-\!{\cal J}_{0}(A_{0}a)]\!=\!0$, subject to the condition $-1\!\leqslant\!\cos(Q_{z}z)
\!=\!\{m_{z}\!+\!2m_{0}[1\!-\!{\cal J}_{0}(A_{0}a)]\}/V_{z}\!\leqslant\!1$. This yields the Weyl-like nodes at
\begin{eqnarray}
z_{c\pm}(A_{0})\!=\!\frac{\lambda_z}{2\pi}\left[\pi\pm\arccos\left(\! \frac{\tilde{m}_{z}(A_{0})}{V_{z}} \!\right) \right],\label{eq:Weyl-like_nodes}
\end{eqnarray} 
where $\tilde{m}_{z}(A_{0})\!=\!m_{z}\!+\!2m_{0}[1\!-\!{\cal J}_{0}(A_{0}a)]$ and $-1\!\leqslant\!\tilde{m}_{z}(A_{0})/V_{z}\!\leqslant\!1$. By substituting the model parameters $m_{z}\!=\!0.5$ eV, $m_{0}\!=\!1$ eV, $V_{z}\!=\!1$ eV, and $a\!=\!1$ nm into the condition $|\tilde{m}_{z}(A_{0})/V_{z}|\!<\!1$, we have $A_{0}\!<\!1.034$ nm$^{-1}$.
From Eq.~\eqref{eq:Weyl-like_nodes}, it is clear that the positions of the Weyl-like nodes can be tuned by adjusting the intensity of the optical field.

Furthermore, substituting Eq.~\eqref{eq:Weyl-like_nodes} into Eq.~\eqref{eq:conductivity3D_last}, we can obtain the total 3D Hall conductivity for the Floquet Weyl-like model in Eq.~\eqref{eq:H_F} as
\begin{eqnarray}
\sigma_{H}^{3D}(A_{0})
\!\approx\! \!\frac{e^{2}\lambda_{z}}{hL_{z}a\pi}\arccos\left(\frac{\tilde{m}_{z}(A_{0})}{V_{z}}\right)\!. \label{eq:conductivity3D_last_F}
\end{eqnarray}

Additionally, Section SVIII in the Supplemental Material~\cite{SuppMat} presents the derivations for the Floquet Weyl-like model under PBCs along the $x$ direction and OBCs along the $y$ direction.

\subsection{Floquet Weyl semimetal}\label{5.2}

The Floquet Hamiltonian for the Weyl semimetal, derived from Eq.~\eqref{eq:HF0}, takes the form
\begin{eqnarray}
{\cal H}^{(F)}_{\rm W}({\bf k})\!=\!\sum_{i=x,y,z}h_{i}\sigma_{i},
\label{eq:H_F_W}
\end{eqnarray} where 
\begin{eqnarray}
h_{x}&\!=\!&{\cal J}_{0}(A_{0}a)t_{x}\sin(k_{x}a) \nonumber\\
&&\!+\!\!\!\!\!\!\!\sum_{n\in{\rm odd},n>0}\!\frac{4t_{x}m_{0}{\cal J}_{n}^{2}(A_{0}a)}{n\hbar\omega}\sin(n\varphi)\sin(k_{x}a)\cos(k_{y}a), \nonumber\\
\\
h_{y}&\!=\!&{\cal J}_{0}(A_{0}a)t_{y}\sin(k_{y}a) \nonumber\\
&&\!+\!\!\!\!\!\!\!\sum_{n\in{\rm odd},n>0}\!\frac{4t_{x}m_{0}{\cal J}_{n}^{2}(A_{0}a)}{n\hbar\omega}\sin(n\varphi)\cos(k_{x}a)\sin(k_{y}a), \nonumber\\
\\
h_{z}&\!=\!&-m_{z}\!+\!t_{z}\cos(k_{z}a)\!-\!2m_{0}[1-{\cal J}_{0}(A_{0}a)] \nonumber\\
&&\!-{\cal J}_{0}(A_{0}a)m_{0}\left[2\!-\!\cos(k_{x}a) \!-\! \cos(k_{y}a)\right] \nonumber\\
&&\!+\!\!\!\!\!\!\!\sum_{n\in{\rm odd},n>0}\!\frac{4t_{x}t_{y}{\cal J}_{n}^{2}(A_{0}a)}{n\hbar\omega}\sin(n\varphi)\cos(k_{x}a)\cos(k_{y}a).\nonumber\\
\end{eqnarray} A detailed derivation of Eq.~\eqref{eq:H_F_W} is provided in Subsection SVII B in the Supplemental Material~\cite{SuppMat}. For linearly polarized light ($\varphi\!=\!0$), the Weyl nodes are determined by solving the equation $-m_{z}\!+\!t_{z}\cos(k_{z}a)\!-\!2m_{0}[1\!-\!{\cal J}_{0}(A_{0}a)]\!=\!0$, subject to the condition $-1\!\leqslant\!\cos(k_{z}a)\!=\!\{m_{z}\!+\!2m_{0}[1\!-\!{\cal J}_{0}(A_{0}a)]\}/t_{z}\!\leqslant\!1$. This condition gives the Weyl node positions
\begin{eqnarray}
k_{c\pm}(A_{0})\!=\!\pm\frac{1}{a}\arccos\left(\! \frac{\tilde{m}_{z}(A_{0})}{t_{z}} \!\right),\label{eq:Weyl_nodes}
\end{eqnarray} where $\tilde{m}_{z}(A_{0})\!=\!m_{z}\!+\!2m_{0}[1\!-\!{\cal J}_{0}(A_{0}a)]$ and $-1\!\leqslant\!\tilde{m}_{z}(A_{0})/t_{z}\!\leqslant\!1$.
Substituting typical model parameters $m_{z}\!=\!0.5$ eV, $m_{0}\!=\!1$ eV, $t_{z}\!=\!1$ eV, and $a\!=\!1$ nm into the condition $|\tilde{m}_{z}(A_{0})/t_{z}|\!<\!1$, we obtain the constraint $A_{0}\!<\!1.034$ nm$^{-1}$.
From Eq.~\eqref{eq:Weyl_nodes}, we can conclude that the positions of the Weyl nodes can be tuned by adjusting the intensity of the optical field.

Furthermore, substituting Eq.~\eqref{eq:Weyl_nodes} into Eq.~\eqref{eq:conductivity3D_Weyl_last}, we can obtain the total 3D Hall conductivity for the Floquet Weyl semimetal in Eq.~\eqref{eq:H_F_W} as
\begin{eqnarray}
\sigma_{H}^{3D}(A_{0})
\!=\!\frac{e^{2}}{ha\pi}\arccos\left(\!\frac{\tilde{m}_{z}(A_{0})}{t_{z}}\!\right). \label{eq:conductivity3D_Weyl_last_F}
\end{eqnarray}

The detailed derivation of the Floquet Hamiltonian matrix for the Weyl semimetal, under OBCs along the $z$ direction and PBCs along the $x$ and $y$ directions, is provided in Subsection SIX A in the Supplemental Material~\cite{SuppMat}. Similarly, Subsection SIX B in the Supplemental Material~\cite{SuppMat} presents the derivation for the case of PBCs along the $x$ and $z$ directions and OBCs along the $y$ direction.

\subsection{Validity}\label{5.3}

To quantitatively assess the validity of the high-frequency expansion, we estimate the maximum instantaneous energy of the time-dependent Hamiltonian ${\cal H}({\bf k},t)\!=\!{\cal H}\left({\bf k} \!-\! \frac{e}{\hbar}{\bf A}(t)\right)$ [Eq.~\eqref{eq:Ht}] averaged over a period of the field. Specifically, we require the condition at the $\Gamma$ point ($k_{x}\!=\!k_{y}\!=\!0$) to be satisfied: $\frac{1}{T}\int_{0}^{T}dt~\text{max}\left\{\big|\big|{\cal H}({\bf k},t)\big|\big|\right\}\!<\!\hbar\omega$. This implies that the optical field parameters must satisfy ${\rm max}\left(t_{x}A_{0}a,t_{y}A_{0}a,m_{0}A_{0}^{2}a^{2}\right)\!<\!\hbar\omega$. In the high-frequency regime, with $\hbar\omega\!=\!10$ eV as an example, we obtain $A_{0}\!<\!{\rm min}\left(\frac{\hbar\omega}{at_x},\frac{\hbar\omega}{at_y}, \frac{1}{a}\sqrt{\frac{\hbar\omega}{m_{0}}}\right)$. For the model parameters $t_{x}\!=\!t_{y}\!=\!1$ eV and $m_{0}\!=\!1$ eV, this gives $A_{0}\!=\!eE_{0}/(\hbar\omega)\!<\!3.16$ (nm)$^{-1}$. Combining the condition $A_{0}\!<\!3.16$ (nm)$^{-1}$ with the constraint for the Weyl nodes $A_{0}<1.034$ (nm)$^{-1}$, we conclude that $A_{0}\!<\!1.034$ (nm)$^{-1}$. This ensures that the high-frequency expansion remains valid under the specified conditions.

\section{Numerical results for Floquet engineering}\label{6}

In this section, we present the numerical results for Floquet engineering, with a comparative analysis between the Floquet Weyl-like and Floquet Weyl semimetal models.

\subsection{Floquet Weyl nodes and total 3D Hall conductivity}\label{6.1}

\begin{figure}
\centering
\includegraphics[width=\columnwidth]{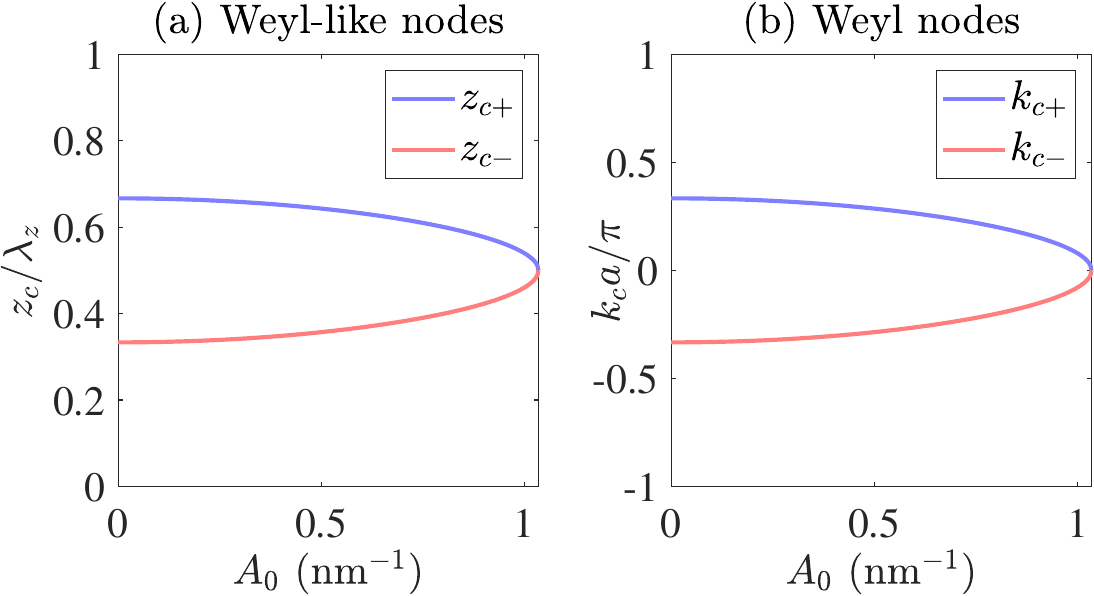}
\caption{(a) Floquet Weyl-like nodes [Eq.~\eqref{eq:Weyl-like_nodes}] and (b) Floquet Weyl nodes [Eq.~\eqref{eq:Weyl_nodes}] as functions of light intensity $A_{0}$. The parameters are $\varphi\!=\!0$ (linearly polarized laser), $m_{z}\!=\!0.5$ eV, $m_{0}\!=\!1$ eV, $t_{x}\!=\!t_{y}\!=\!1$ eV, $V_{z}\!=\!1$ eV, $t_{z}\!=\!1$ eV, and $L_{z}\!=\!\lambda_z\!=\!31a$.}
\label{fig_Weyl_nodes_linearly_together}
\end{figure}

Based on Eqs.~\eqref{eq:Weyl-like_nodes} and \eqref{eq:Weyl_nodes}, the positions of the Floquet Weyl-like nodes and Floquet Weyl nodes can be plotted, as shown in Fig.~\ref{fig_Weyl_nodes_linearly_together}. Figures~\ref{fig_Weyl_nodes_linearly_together}(a) and \ref{fig_Weyl_nodes_linearly_together}(b) illustrate that the locations of these nodes are tunable via the light intensity $A_{0}$. Notably, when $A_{0}$ reaches a critical value of $\sim1.034$ (nm)$^{-1}$, the two Floquet Weyl-like nodes (or Floquet Weyl nodes) merge into a single point, indicating a topological phase transition.

\begin{figure}
\centering
\includegraphics[width=\columnwidth]{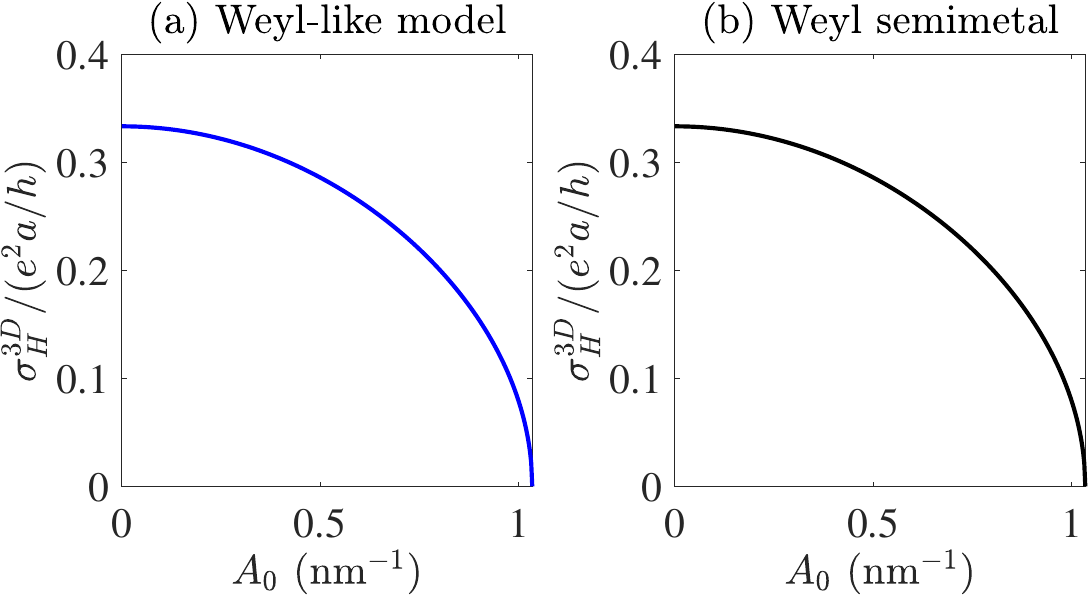}
\caption{Total 3D Hall conductivities calculated from Eqs.~\eqref{eq:conductivity3D_last_F} and \eqref{eq:conductivity3D_Weyl_last_F} for (a) the Floquet Weyl-like model and (b) the Floquet Weyl semimetal, as functions of light intensity $A_{0}$. All other parameters are the same as those in Fig.~\ref{fig_Weyl_nodes_linearly_together}.}
\label{fig_Hall3D_linearly_together}
\end{figure}

In parallel, using Eqs.~\eqref{eq:conductivity3D_last_F} and \eqref{eq:conductivity3D_Weyl_last_F}, the total 3D Hall conductivity as a function of $A_{0}$ is calculated for both the Floquet Weyl-like model and the Floquet Weyl semimetal, as shown in Fig.~\ref{fig_Hall3D_linearly_together}. The results show that the total 3D Hall conductivity decreases monotonically with increasing $A_{0}$, ultimately approaching zero as $A_{0}$ nears the critical value of $\sim1.034$ (nm)$^{-1}$.

\subsection{Floquet Chern number}\label{6.2}

\begin{figure}
\centering
\includegraphics[width=\columnwidth]{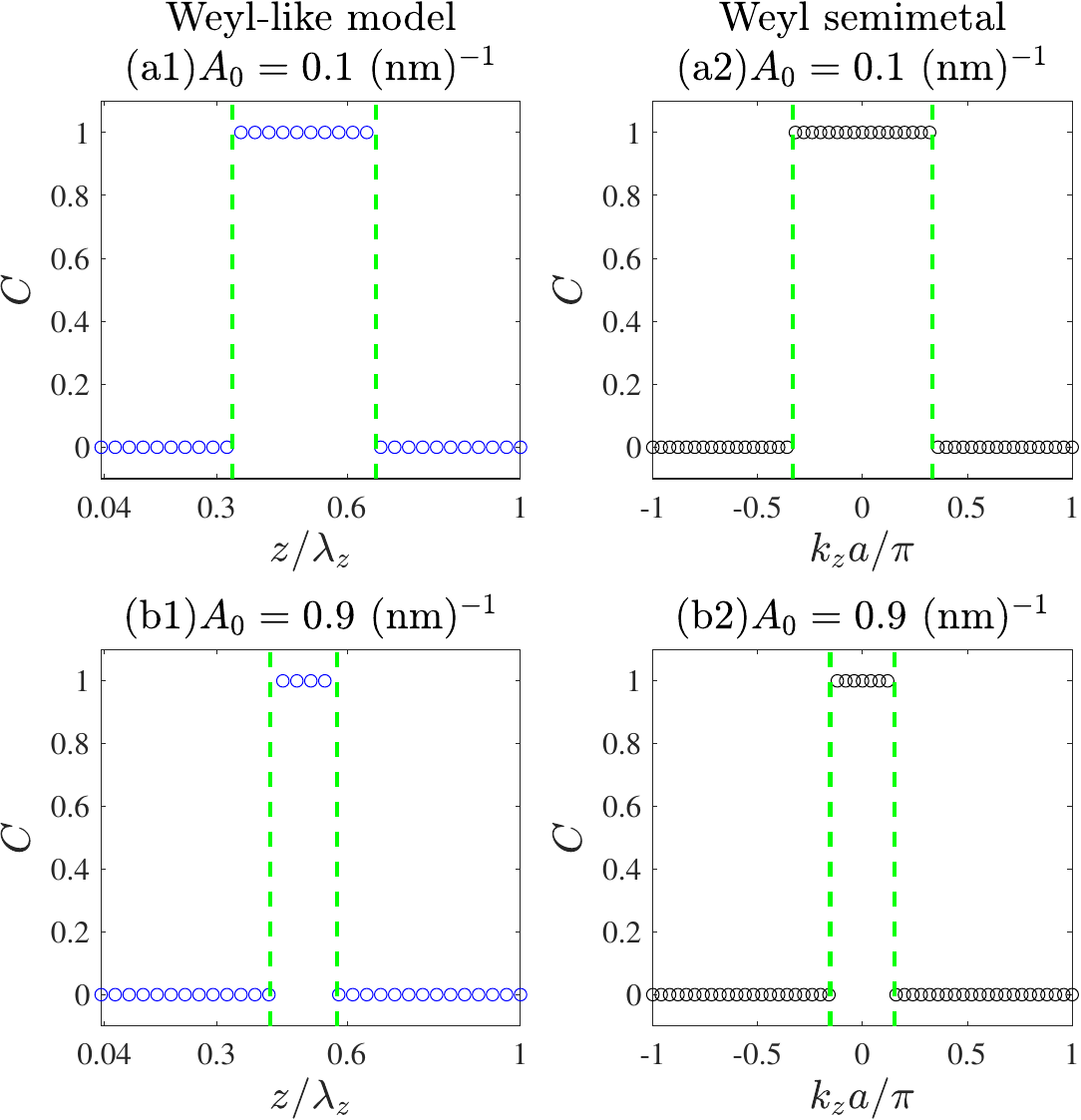}
\caption{Chern numbers for (a1) and (b1) the Floquet Weyl-like model and (a2) and (b2) the Floquet Weyl semimetal. Panels (a1) and (a2) correspond to $A_{0} \!=\! 0.1$ (nm)$^{-1}$, and panels (b1) and (b2) correspond to $A_{0} \!=\! 0.9$ (nm)$^{-1}$. The two green vertical dashed lines indicate the positions of the Weyl-like or Weyl points. All other parameters are the same as those in Fig.~\ref{fig_Weyl_nodes_linearly_together}.}
\label{fig_C_Dual_linearly}
\end{figure}

We compute the Chern numbers for both the Floquet Weyl-like model and the Floquet Weyl semimetal model, as shown in Fig.~\ref{fig_C_Dual_linearly}. The results indicate that the phase transition points, represented by the Floquet Weyl nodes, shift with varying light intensities. Detailed derivations of the Chern numbers for both Floquet models are provided in Sections SX and SXI in the Supplemental Material~\cite{SuppMat}, respectively.

\subsection{Floquet energy bands}\label{6.3}

\begin{figure}
\centering
\includegraphics[width=\columnwidth]{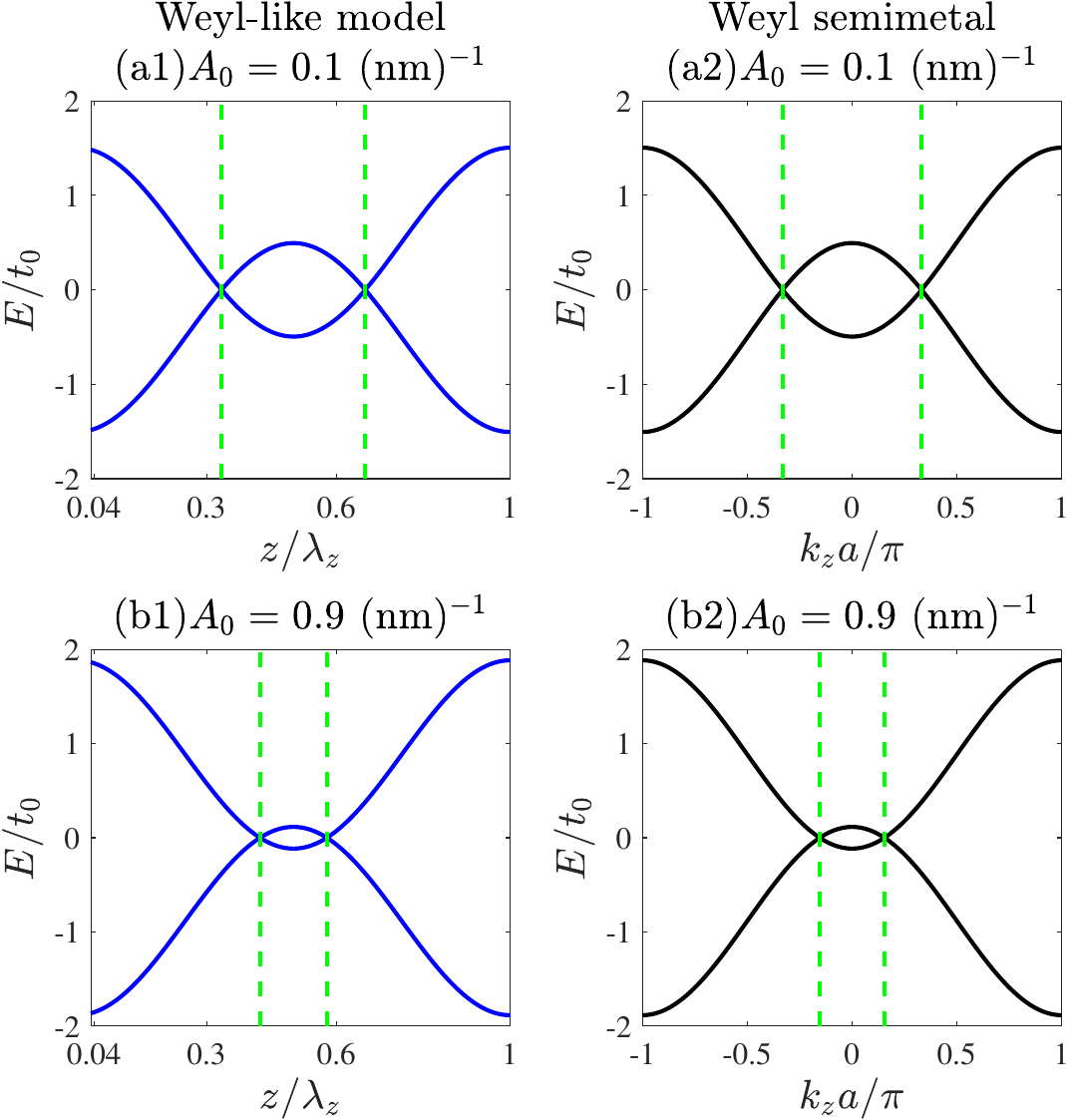}
\caption{Energy bands for (a1) and (b1) the Floquet Weyl-like model and (a2) and (b2) the Floquet Weyl semimetal with $k_x \!=\! k_y \!=\! 0$. Panels (a1) and (a2) correspond to $A_{0} \!=\! 0.1$ (nm)$^{-1}$, and panels (b1) and (b2) to $A_{0} \!=\! 0.9$ (nm)$^{-1}$. The two green vertical dashed lines mark the positions of the Weyl-like or Weyl points. All other parameters are the same as those in Fig.~\ref{fig_Weyl_nodes_linearly_together}.}
\label{fig_E_Dual_linearly}
\end{figure}

The energy bands for the Floquet Weyl-like model and the Floquet Weyl semimetal, calculated at $k_x\!=\!k_y\!=\!0$, are shown in Fig.~\ref{fig_E_Dual_linearly}. The two green vertical dashed lines are used to label the positions of the two Weyl-like (or Weyl) points. The energy gapless points, corresponding to the Weyl nodes, exhibit a dependence on the light intensity.

\subsection{Floquet Fermi arc}\label{6.4}

\begin{figure}
\centering
\includegraphics[width=\columnwidth]{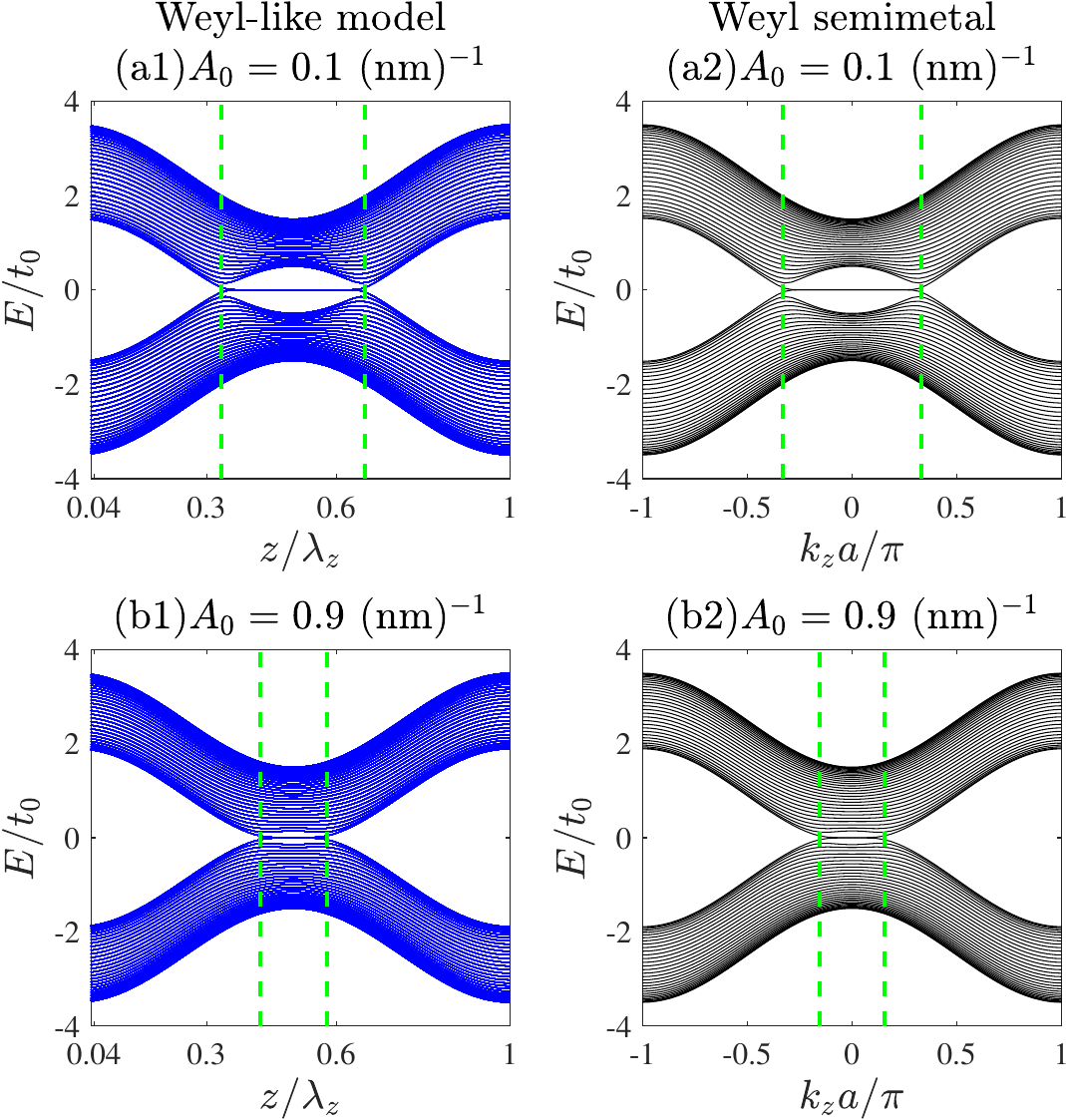}
\caption{Energy bands (solid curves) and Fermi arcs (horizontal line at $E \!=\! 0$) for (a1) and (b1) the Floquet Weyl-like model and (a2) and (b2) the Floquet Weyl semimetal under open boundary conditions (OBCs) along the $y$ direction, with $k_x \!=\! 0$ and $L_y \!=\! 31a$. Panels (a1) and (a2) correspond to $A_{0} \!=\! 0.1$ (nm)$^{-1}$ and panels (b1) and (b2) to $A_{0} \!=\! 0.9$ (nm)$^{-1}$. The two green vertical dashed lines indicate the positions of the Weyl-like or Weyl points. All other parameters are the same as those in Fig.~\ref{fig_Weyl_nodes_linearly_together}.}
\label{fig_edge_E_Dual_linearly}
\end{figure}

In Fig.~\ref{fig_edge_E_Dual_linearly}, we show the energy bands (solid curves) and Fermi arcs (depicted as horizontal straight lines at $E\!=\!0$) for both the Floquet Weyl-like model and the Floquet Weyl semimetal under OBCs along the $y$ direction, with $k_x\!=\!0$ and $L_y\!=\!31a$. We observe that the lengths of the Floquet Fermi arcs change with light intensity, reflecting variations in the Floquet Weyl nodes.

\section{Experimental Realizations}\label{7}

In this section, we discuss potential experimental realizations of the Weyl-like model. Weyl semimetals have been successfully realized in various platforms, including electrical circuits~\cite{lu2019probing,lee2020imaging}, ultracold atomic gases~\cite{wang2021realization,lu2020ideal}, photonic crystals~\cite{lu2015experimental,li2018weyl}, and acoustic metamaterials~\cite{xiao2015synthetic,yang2016acoustic,peri2019axial}. For the purpose of illustration, we focus on the realization of the Weyl-like model using electrical circuits.

Unlike the 3D cubic lattices employed in cold-atom proposals~\cite{dubvcek2015weyl,roy2018tunable} and experimental realizations~\cite{wang2021realization,lu2020ideal}, we propose a design based on a stack of 2D square lattices, with each plaquette in every principal plane hosting a flux of $\pi$, as described in Ref.~\cite{ningyuan2015time}. This configuration corresponds to a system of spin-$\frac{1}{2}$ particles with engineered spin-dependent tunneling.

The system thus takes the form of a multilayer 2D time-reversal symmetric tight-binding model, with two sites per unit cell. This differs from the 3D time-reversal symmetric tight-binding model on a cubic lattice, which was originally proposed in Ref.~\cite{lu2019probing} to realize Weyl semimetals in 3D momentum space.

To implement this multilayer 2D square lattice in an electrical circuit, each lattice site is represented by an inductor, while tunnel couplings are modeled by pairs of capacitors. Zero-phase tunnel couplings connect the positive ends of adjacent inductors, while their negative ends are similarly linked. A $\pi$-phase tunnel coupling capacitively links the positive end of one inductor to the negative end of a neighboring inductor and vice versa.

The stacking of the 2D square lattice layers is coupled with a spatially periodic potential, which can be induced by an alternating current~\cite{qin2024light} or via the charge-density-wave mechanism~\cite{zhang2023emergent,qin2020theory}. This approach offers a promising platform for simulating the Weyl-like model, enabling experimental exploration of its topological properties.

\section{Realization for both type-I and type-II Weyl-like model behaviors}\label{8} 

\subsection{Continuous Model}\label{8.1} 

To investigate the emergence of both type-I and II Weyl-like behaviors, we consider a tilted continuous Weyl-like model Hamiltonian in the $(k_x, k_y, z)$ space, given by
\begin{eqnarray}
{\cal H}({\bf k}_{||},z)
&\!=\!&\left\{-M_{0}\!+\!C_{z}(Q_{z}z)^{2}\!+\!C_{||}[(k_{x}a)^{2}\!+\!(k_{y}a)^{2}]\right\}\!\sigma_{0}^{} \nonumber\\
&&\!+\left\{M_{z}\!-\!J_{z}(Q_{z}z)^{2}\!-\!J_{||}[(k_{x}a)^{2}\!+\!(k_{y}a)^{2}]\right\}\!\sigma_{z} \nonumber\\
&&\!+v_{0}^{}k_{x}a\sigma_{x}\!+\!v_{0}^{}k_{y}a\sigma_{y}, \label{eq:H_z_typeII}
\end{eqnarray} where $Q_{z}$ is the wave vector of the spatially periodic potential along the $z$ direction; ${\bf k}_{||}\!=\!(k_x, k_y)$; $a$ is the lattice constant; $\sigma_{0}^{}$ is a $2\times2$ identity matrix; $M_{0}$, $C_{z}$, $C_{||}$, $M_{z}$, $J_{z}$, $J_{||}$, and $v_{0}^{}$ are model parameters. 
The corresponding eigenenergies are
\begin{eqnarray}
&&E_{\pm}({\bf k}_{||},z) \nonumber\\
&\!=\!&-M_{0}\!+\!C_{z}(Q_{z}z)^{2}\!+\!C_{||}[(k_{x}a)^{2}\!+\!(k_{y}a)^{2}] \nonumber\\
&&\pm\sqrt{(v_{0}^{}a)^{2}(k_{x}^{2}\!+\!k_{y}^{2})\!+\!\left[M_{z}\!-\!J_{z}(Q_{z}z)^{2}\!-\!J_{||}a^{2}(k_{x}^{2}\!+\!k_{y}^{2})\right]^{2}},\label{eq:E_z0_typeII}\nonumber\\
\end{eqnarray} 
which indicates two Weyl-like points at $z_{c\pm}\!=\!\pm (1/Q_{z})\sqrt{M_{z}/J_{z}}$ when $k_{x}\!=\!k_{y}\!=\!0$.

\begin{figure}
\centering
\includegraphics[width=\columnwidth]{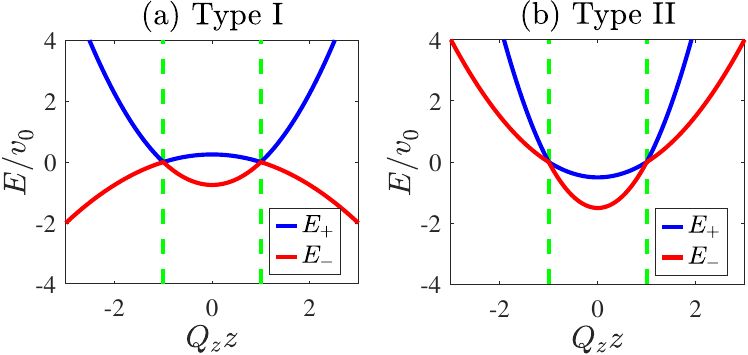}
\caption{Energy bands of the tilted Weyl-like model with $k_{x}\!=\!k_{y}\!=\!0$ for (a) type I ($C_{z}\!=\!0.25$ eV $\!<\!$ $J_{z}\!=\!0.5$ eV) and (b) type II ($C_{z}\!=\!1$ eV $\!>\!$ $J_{z}\!=\!0.5$ eV). Vertical dashed green lines indicate the Weyl-like points. Other parameters: $M_{z}\!=\!0.5$ eV, $M_{0}\!=\!C_{z}M_{z}/J_{z}$, $v_{0}^{}\!=\!1$ eV, and $a\!=\!1$ nm.}
\label{fig_WLP_I_II_E}
\end{figure}

Expanding Eq.~\eqref{eq:E_z0_typeII} around the Weyl-like points $(0,0,z_{c\pm})$~\cite{chen2018floquet1,qin2020theory}, we obtain the linearized energy bands as
\begin{eqnarray}
E_{\pm}(z)&\!\approx\!&-M_{0}\!+\!C_{z}(Q_{z}z_{c\pm})^{2}\!-\!2C_{z}Q_{z}^{2}z_{c\pm}z \nonumber\\
&&\!\pm \Big|M_{z}\!-\!J_{z}(Q_{z}z_{c\pm})^{2}\!+\!2J_{z}Q_{z}^{2}z_{c\pm}z\Big| \nonumber\\
&\!=\!&-M_{0}\!+\!C_{z}\frac{M_{z}}{J_{z}}\!\mp\!2C_{z}\sqrt{\frac{M_{z}}{J_{z}}}Q_{z}z \pm \Big|2\sqrt{M_{z}J_{z}}Q_{z}z\Big| \nonumber\\
&\!=\!&\mp2C_{z}\sqrt{\frac{M_{z}}{J_{z}}}Q_{z}z \pm \Big|2\sqrt{M_{z}J_{z}}Q_{z}z\Big|,\label{eq:E_z0_typeII_linear} 
\end{eqnarray} where we used $(z\!-\!z_{c\pm})^{2}\!\approx\!z_{c\pm}^{2}\!-\!2z_{c\pm}z$ and assumed $M_{0}\!=\!C_{z}M_{z}/J_{z}$ for simplicity.
The first term $\mp 2 C_z \sqrt{M_z / J_z} Q_z z$ characterizes the tilt of the energy bands, while the second term $\pm 2 \sqrt{M_z J_z} Q_z |z|$ describes the band splitting. When the tilt dominates the splitting, i.e., $\Big|2C_{z}\sqrt{M_{z}/J_{z}}Q_{z}z\Big| \!>\! \Big|2\sqrt{M_{z}J_{z}}Q_{z}z\Big|$, the Weyl-like point becomes type II.
We define a tilt ratio as~\cite{chen2018floquet1,soluyanov2015type,li2017type}
\begin{eqnarray}
F_{z}
\!=\!\frac{\Big|2C_{z}\sqrt{M_{z}/J_{z}}Q_{z}z\Big|}{\Big|2\sqrt{M_{z}J_{z}}Q_{z}z\Big|}
\!=\!\frac{C_{z}}{J_{z}}. \label{eq:tilt_ratio_z_typeII}
\end{eqnarray}
Thus, for $F_{z}\!<\!1$ (i.e., $C_{z}\!<\!J_{z}$), the tilt is weak, resulting in a type-I Weyl-like point [Fig.~\ref{fig_WLP_I_II_E}(a)]. Conversely, for $F_{z}\!>\!1$ (i.e., $C_{z}\!>\!J_{z}$), the tilt is strong and both bands radiate in the same direction, indicating a type-II Weyl-like point [Fig.~\ref{fig_WLP_I_II_E}(b)]. Therefore, by tuning $C_{z}$ while keeping other parameters fixed, both type-I and II Weyl-like behaviors can be realized.

As a dual picture, the tilted continuous Weyl semimetal model is given by
\begin{eqnarray}
{\cal H}_{\rm W}({\bf k})
&\!=\!&\left\{-M_{0}\!+\!C_{z}(k_{z}a)^{2}\!+\!C_{||}[(k_{x}a)^{2}\!+\!(k_{y}a)^{2}]\right\}\!\sigma_{0} \nonumber\\
&&\!+\left\{M_{z}\!-\!J_{z}(k_{z}a)^{2}\!-\!J_{||}[(k_{x}a)^{2}\!+\!(k_{y}a)^{2}]\right\}\!\sigma_{z} \nonumber\\
&&\!+v_{0}^{}k_{x}a\sigma_{x}\!+\!v_{0}^{}k_{y}a\sigma_{y}, \label{eq:H_kz_Weyl_typeII}
\end{eqnarray} where ${\bf k}\!=\!(k_x,k_y,k_z)$, and we have replaced $Q_{z}z$ by $k_{z}a$ in Eq.~\eqref{eq:H_z_typeII}. Similar criteria apply here, allowing one to tune between type-I and II Weyl semimetal phases by adjusting $C_{z}$ and $J_{z}$.

\subsection{Berry curvature and 2D Hall conductance}\label{8.2} 

For the tilted continuous Weyl-like model in Eq.~\eqref{eq:H_z_typeII}, the layer Berry curvature is given by~\cite{kubo1957statistical_1,kubo1957statistical_2,thouless1982quantized,shen2017topological,qin2023light,qin2024kinked,qin2024light,qin2025nonlinear}
\begin{eqnarray}
&&{\cal F}_{xy}^{(\pm)}({\bf k}_{||},z) \nonumber\\
&\!=\!&-2\varepsilon_{xyz}\!\frac{{\rm Im}\!\left\langle \!\psi_{\pm}\!\left|\!\frac{\partial{\cal H}}{\partial k_{x}}\right|\!\psi_{\mp}\!\right\rangle\!\!\left\langle\!\psi_{\mp}\!\left|\!\frac{\partial\!{\cal H}}{\partial k_{y}}\!\right|\!\psi_{\pm}\!\right\rangle}{\left(E_{\pm}-E_{\mp}\right)^{2}} \nonumber\\
&\!=\!& \mp\frac{v_{0}^{2}a^{2}\left[M_{z}\!-\!J_{z}(Q_{z}z)^{2}\!+\!J_{||}a^{2}k_{||}^{2}\right]}{2\left\{ v_{0}^{2}a^{2}k_{||}^{2}\!+\!\left[M_{z}\!-\!J_{z}(Q_{z}z)^{2}\!-\!J_{||}a^{2}k_{||}^{2}\right]^{2} \right\}^{3/2}}, \label{eq:Berry_z_typeII} \nonumber\\
\end{eqnarray} where $k_{||}^{2}\!=\!k_{x}^{2}\!+\!k_{y}^{2}$, $\varepsilon_{xyz}$ is the Levi-Civita antisymmetric tensor, ${\cal H}$ is the model Hamiltonian, and $|\psi_{\pm}\rangle$ are the eigenvectors of the energy bands $E_{\pm}$.

To obtain a quantized 2D Hall conductance, we restrict attention to the type-I regime ($C_{z}\!<\!J_{z}$), where the Fermi level $E_{F}\!=\!0$ lies in the band gap, as shown in Fig.~\ref{fig_WLP_I_II_E}(a).

The $z$-dependent 2D Hall conductance for each layer is~\cite{qi2006topological,lu2010massive,shan2010effective,qin2022phase,chen2025probing}
\begin{eqnarray}
\sigma_{H(\pm)}^{2D}(z)&\!=\!&\frac{e^{2}}{h}\frac{1}{2\pi}\int{\cal F}_{xy}^{(\pm)}({\bf k}_{||},z)d^{2}{\bf k}_{||} \nonumber\\
&\!=\!& \pm\frac{e^{2}}{2h}\frac{1}{2\pi}\int_{0}^{2\pi}d\phi\int_{0}^{\infty}dk_{||}^{2}\frac{\partial \cos\theta}{\partial k_{||}^{2}} \nonumber\\
&\!=\!& \mp\frac{e^{2}}{2h}\Big[\text{sgn}(J_{||}) \!+\! \text{sgn}(M_{z}\!-\!J_{z}(Q_{z}z)^{2}) \Big]\!,\label{eq:Chern_z_typeII}\nonumber\\
\end{eqnarray} where $\text{sgn}(x)$ denotes the sign function, which indicates the sign of the variable $x$, and 
\begin{eqnarray}
\cos\theta \!=\! \frac{M_{z}\!-\!J_{z}(Q_{z}z)^{2}\!-\!J_{||}a^{2}k_{||}^{2}}{\left\{ v_{0}^{2}a^{2}k_{||}^{2}\!+\!\left[M_{z}\!-\!J_{z}(Q_{z}z)^{2}\!-\!J_{||}a^{2}k_{||}^{2}\right]^{2} \right\}^{1/2}}.\nonumber\\
\end{eqnarray} 
Therefore, the layer Chern number is given by
\begin{eqnarray}\label{eq:Chern_number_z}
C_{\pm}(z) \!=\! \mp\frac{1}{2}\Big[\text{sgn}(J_{||}) \!+\! \text{sgn}(M_{z}\!-\!J_{z}(Q_{z}z)^{2}) \Big]\!.
\end{eqnarray}

Similarly, for the tilted continuous Weyl semimetal model in Eq.~\eqref{eq:H_kz_Weyl_typeII}, the Berry curvature reads
\begin{eqnarray}
{\cal F}_{xy}^{(\pm)}({\bf k}) 
\!=\!\mp\frac{v_{0}^{2}a^{2}\left[M_{z}\!-\!J_{z}a^{2}k_z^{2}\!+\!J_{||}a^{2}k_{||}^{2}\right]}{2\left\{ v_{0}^{2}a^{2}k_{||}^{2}\!+\!\left[M_{z}\!-\!J_{z}a^{2}k_z^{2}\!-\!J_{||}a^{2}k_{||}^{2}\right]^{2} \right\}^{3/2}}. \label{eq:Berry_kz_Weyl_typeII} \nonumber\\
\end{eqnarray} 
Furthermore, the corresponding 2D Hall conductance is given by 
\begin{eqnarray}
\sigma_{H(\pm)}^{2D}(k_z)&\!=\!&\frac{e^{2}}{h}\frac{1}{2\pi}\int{\cal F}_{xy}^{(\pm)}({\bf k}_{||},k_z)d^{2}{\bf k}_{||} \nonumber\\
&\!=\!& \mp\frac{e^{2}}{2h}\Big[\text{sgn}(J_{||}) \!+\! \text{sgn}(M_{z}\!-\!J_{z}(k_{z}a)^{2}) \Big]\!,\label{eq:Chern_kz_Weyl_typeII} \nonumber\\ 
\end{eqnarray}
and the corresponding Chern number becomes 
\begin{eqnarray}\label{eq:Chern_number_kz}
C_{\pm}(k_{z}) \!=\! \mp\frac{1}{2}\Big[\text{sgn}(J_{||}) \!+\! \text{sgn}(M_{z}\!-\!J_{z}(k_{z}a)^{2}) \Big]\!.
\end{eqnarray}

\subsection{3D Hall conductivity}\label{8.3} 

The 3D Hall conductivity for the tilted type-I Weyl-like model in Eq.~\eqref{eq:H_z_typeII} is computed as
\begin{eqnarray}
\sigma_{H(\pm)}^{3D}
&\!=\!&\sum_{j_z=-L_{z}/(2a)}^{L_{z}/(2a)}\frac{\sigma_{H(\pm)}^{2D}(z)}{L_{z}}
\!=\!\frac{e^{2}}{h}\sum_{j_z=z_{c-}/a}^{z_{c+}/a}\frac{C_{\pm}(z)}{L_{z}} \nonumber\\
&\!\approx\!& \mp\frac{2e^{2}}{hL_{z}aQ_{z}}\sqrt{\frac{M_{z}}{J_{z}}}\!, \label{eq:conductivity3D_typeII}
\end{eqnarray} where $j_{z}\!=\!z/a$ is the layer index, $L_z$ is the system length along the $z$ direction, and we used $z_{c\pm}\!=\!\pm (1/Q_{z})\sqrt{M_{z}/J_{z}}$, and the layer Chern number $C_{\pm}(z)$ [Eq.~\eqref{eq:Chern_number_z}] satisfies
\begin{eqnarray}
C_{\pm}(z)&\!=\!&\!\left\{ \begin{array}{l}
\mp1,~~z_{c-}\!<\!z\!<\!z_{c+}~{\rm and}~J_{||}\!>\!0; \\
0,~~z\!<\!z_{c-}~{\rm or}~z\!>\!z_{c+}~{\rm and}~J_{||}\!>\!0.
\end{array}\right.
\end{eqnarray} Similarly, the 3D Hall conductivity of the type-II Weyl-like model also depends on $L_z$, owing to the summation over the layer index $j_z\!=\!z/a$ in its definition for the Weyl-like model. 

For the tilted continuous type-I Weyl semimetal model in Eq.~\eqref{eq:H_kz_Weyl_typeII}, the 3D Hall conductivity is given by
\begin{eqnarray}
\sigma_{H(\pm)}^{3D}
&\!=\!&\int_{-\pi/a}^{\pi/a}\frac{dk_{z}}{2\pi}\sigma_{H(\pm)}^{2D}(k_z)\!=\!\frac{e^{2}}{h}\int_{k_{c-}}^{k_{c+}}C_{\pm}(k_z)\frac{dk_{z}}{2\pi}  \nonumber\\
&\!=\!&\mp\frac{e^{2}}{ha\pi}\sqrt{\frac{M_{z}}{J_{z}}}, \label{eq:conductivity3D_Weyl_typeII}
\end{eqnarray} where we used $k_{c\pm}\!=\!\pm(1/a)\sqrt{M_{z}/J_{z}}$, and the Chern number $C_{\pm}(k_{z})$ [Eq.~\eqref{eq:Chern_number_kz}] satisfies
\begin{eqnarray}
C_{\pm}(k_{z})&\!=\!&\!\left\{ \begin{array}{l}
\mp1,~~k_{c-}\!<\!k_{z}\!<\!k_{c+}~{\rm and}~J_{||}\!>\!0; \\
0,~~k_{z}\!<\!k_{c-}~{\rm or}~k_{z}\!>\!k_{c+}~{\rm and}~J_{||}\!>\!0.
\end{array}\right.
\end{eqnarray}

By comparing Eqs.~\eqref{eq:conductivity3D_typeII} and \eqref{eq:conductivity3D_Weyl_typeII}, we find that the 3D Hall conductivity of the Weyl-like model explicitly depends on the system size $L_z$, while that of the Weyl semimetal is independent of system size. This finite-size dependence provides a clear and unique signature of mixed-space topology in the Weyl-like model.

\section{Summary}\label{9}

In our study, we introduce a 3D topological phase with Weyl-semimetal-like characteristics, which is induced by a periodic potential. This phase is realized by stacking 2D Chern insulators, leading to Weyl-like points in the parameter space $(k_x, k_y, z)$, in contrast with the conventional Weyl points in momentum space $(k_x, k_y, k_z)$.

The Weyl-semimetal-like phase we propose shares key features with traditional Weyl semimetals, such as linear dispersion near the Weyl-like points, nontrivial bulk topology, the emergence of Fermi arc that connect these points, and the Berry monopoles. Importantly, these topological behaviors manifest in real space rather than in momentum space.

We further analyze the role of interlayer coupling by computing the local density of states, the layer Hall conductance, and the total 3D Hall conductivity and show that the Weyl-semimetal-like phase remains robust under weak and moderate coupling. The effects of disorder are also investigated: Beyond a critical disorder strength, the Weyl-like points destabilize, driving the system into a trivial phase.

Additionally, we explore Floquet engineering under high-frequency periodic driving, calculating Floquet Weyl nodes, Floquet Chern numbers, Floquet energy bands, and Floquet Fermi arcs. Our results demonstrate that the Weyl-like nodes can be effectively tuned by laser fields, offering a route to manipulate topological properties dynamically.

Finally, we reveal the emergence of both type-I and II Weyl-like behaviors in a tilted Weyl-like model.
In this work, we open possibilities for exploring and controlling topological phases through external stimuli.

\begin{acknowledgments}
We acknowledge helpful discussions with Ching Hua Lee. F.Q. acknowledges support from the Jiangsu Specially-Appointed Professor Program in Jiangsu Province and the Doctoral Research Start-Up Fund of Jiangsu University of Science and Technology. R.C. acknowledges support from the National Natural Science Foundation of China (Grant No. 12304195), the Chutian Scholars Program in Hubei Province, the Hubei Provincial Natural Science Foundation (Grant No. 2025AFA081), and the Original Seed Program of Hubei University.
\end{acknowledgments}

\section*{Data Availability}
The data are available from the authors upon reasonable request.

\bibliography{references_WSM}

\end{document}